\documentclass[sigconf]{acmart}

\AtBeginDocument{%
  }
    
\usepackage{subcaption}
\usepackage{url}

\usepackage{acmart-taps}
\usepackage{float}  %
\usepackage{stfloats}

\usepackage{setspace}

\usepackage{fancybox} 

\makeatletter 
\newcommand{\OvalColumnBox}[1]{%
\par\medskip
\begingroup
\cornersize{.2}
\setlength{\fboxsep}{8pt}
\setlength{\fboxrule}{0.8pt}
\noindent\makebox[\columnwidth][l]{%
\ovalbox{%
\begin{minipage}{\dimexpr\columnwidth-2\fboxsep-2\fboxrule\relax}%
\par\medskip
#1
\end{minipage}%
}%
}%
\endgroup
\par\medskip
}
\makeatother

\usepackage{siunitx}

\usepackage{fancyvrb}

\usepackage{microtype}

\copyrightyear{2026}
\acmYear{2026}
\setcopyright{usgovmixed}
\acmConference[CHI '26]{Proceedings of the 2026 CHI Conference on Human Factors in Computing Systems}{April 13--17, 2026}{Barcelona, Spain}
\acmBooktitle{Proceedings of the 2026 CHI Conference on Human Factors in Computing Systems (CHI '26), April 13--17, 2026, Barcelona, Spain}
\acmDOI{10.1145/3772318.3791428}
\acmISBN{979-8-4007-2278-3/2026/04}
\begin{document}

\title{Who Does What? Archetypes of Roles Assigned to LLMs During Human-AI Decision-Making}

\author{Shreya Chappidi}
\affiliation{%
  \institution{University of Cambridge}
  \city{Cambridge}
  \country{United Kingdom}
}
\affiliation{%
  \institution{National Cancer Institute, National Institutes of Health}
  \city{Bethesda, MD}
  \country{United States}
}

\author{Jatinder Singh}
\affiliation{%
  \institution{Research Centre Trust, UA Ruhr, University Duisburg-Essen}
  \city{Duisburg}
  \country{Germany}
}
\affiliation{%
  \institution{University of Cambridge}
  \city{Cambridge}
  \country{United Kingdom}}

\author{Andra V. Krauze}
\affiliation{%
  \institution{National Cancer Institute, National Institutes of Health}
  \city{Bethesda, MD}
  \country{United States}
}

\renewcommand{\shortauthors}{Chappidi et al.}

\begin{abstract}

LLMs are increasingly supporting decision-making across high-stakes domains, requiring critical reflection on the socio-technical factors that shape how humans and LLMs are assigned roles and interact during human-in-the-loop decision-making. 
{This paper introduces the concept of \textit{human-LLM archetypes} -- defined as recurring socio-technical interaction patterns} that structure the roles of humans and LLMs in collaborative decision-making.
{We describe 17 human-LLM archetypes derived from a scoping literature review and thematic analysis of 113 LLM-supported decision-making papers.}
Then, we evaluate these diverse archetypes across real-world clinical diagnostic cases to examine the potential effects of adopting distinct human-LLM archetypes on LLM outputs and decision outcomes. 
Finally, we present relevant tradeoffs and design choices across human-LLM archetypes, including decision control, social hierarchies, cognitive forcing strategies, and information requirements.
Through our analysis, we show that selection of human-LLM interaction archetype can influence LLM outputs and decisions, bringing important risks and considerations for the designers of human-AI decision-making systems.

\end{abstract}



\maketitle

\section{Introduction}
\label{sec:introduction}

Large language models (LLMs) are widely  being applied to various problem-solving and decision-making tasks, including finance, web search, medicine, and logistics, where they detect patterns and communicate findings in natural language. As a result, LLMs are being assigned increasingly complex socio-technical {``roles''} (e.g., fact-checker, information-seeking agent, arbiter, etc.), expanding their influence on everyday decisions. %

Most LLM evaluations use question-answering\slash benchmark tasks. While useful for measuring isolated capabilities such as reasoning or information retrieval, this approach overlooks how LLMs are deployed in practice -- with multi-faceted, extended interactions embedded in \textit{human-in-the loop} (HITL) workflows \cite{miller_evaluating_2025, griot_large_2025, fodor_line_2025, mcintosh_inadequacies_2025}. At the same time, human-AI (HAI) %
researchers are increasingly advocating that LLMs should be designed to augment human capabilities and workflows, not merely replicate human skills or perform static knowledge checks \cite{sokol_artificial_2025}.

HITL workflows involve role, contextual, and temporal dynamics that can shape performance and decision-making outcomes, yet these factors are often unconsidered during LLM evaluations. For example, multiple design choices shape human-LLM interactions during decision-making, including model pretraining and fine-tuning, input/output format (e.g., single-shot question answering, chain-of-thought, or chatbot interfaces) \cite{wei_chain--thought_2023}, prompt structure and syntax \cite{gourabathina_medium_2025}, system messages \cite{neumann_position_2025}, and the ordering of decision-workflow steps \cite{tamkin_evaluating_2023}. 

At the same time, concerns related to LLM hallucination \cite{huang_survey_2025}, overreliance \cite{ashktorab_emerging_2025, kim_fostering_2025, bo_rely_2025}, sycophancy \cite{malmqvist_sycophancy_2024}, and various biases \cite{kuhl_when_2025, sun_sociodemographic_2025} have led to critiques and calls for caution over real-world deployment. As organizations across sectors increasingly adopt LLMs, it will be important to understand and interrogate how design choices within human-LLM interactions, including problem formulation and LLM role assignment, affect high-stakes decision-making.

Towards this, in this paper we introduce the concept of \textbf{human-LLM \textit{archetypes}---distinct socio-technical patterns of human-LLM interaction in decision-making}---which capture collaborative dynamics, problem formulation and framing, cognitive workflows, and related HCI dimensions. While system details may vary between individual implementations, this {conception}
of human-LLM archetypes allows us to identify, compare, and contrast the broader factors that determine how LLMs are employed and integrated into real-world decision workflows. Employing this framework to analyze prototypical approaches to human-LLM decision-making also enables us to highlight design choices and potential risks requiring additional consideration by designers and users of human-LLM systems. {In all, our archetypes framework enables deeper understandings of the socio-technical factors governing human-LLM roles and interactions, can be flexibly applied across decision-making domains, contexts, and goals, and can guide the responsible design and deployment of human-LLM decision-making systems.}

\subsection{Contributions}
We consider the socio-technical {\textit{roles} of LLMs---defined here as the positions or functions assigned to an LLM during human-LLM decision-making interactions}.
The roles assigned to LLMs and associated system design within a human-AI decision-making-framework will shape the deliberation process and associated outcome.
While there are emerging critical studies observing disparate LLM outputs across prompting strategies \cite{sun_sociodemographic_2025, wan_are_2023, gourabathina_medium_2025}, they are often limited to scenarios devised for evaluation purposes and do not often engage or discuss the human- or cognitive-centered aspects of how the LLM is deployed. %

Our contributions include: 
\begin{enumerate}
    \item 
    A framework that systematically \textbf{taxonomizes human-LLM archetypes}---recurring socio-technical interaction patterns that structure the roles assigned to LLMs during human-in-the-loop decision-making. This involves identifying critical yet underconsidered factors of problem framing, human-LLM task assignments, cognitive load, and social positioning.
    \item An empirical case study evaluation of the \textbf{potential impacts of adopting different LLM archetypes} by comparing LLM outputs across factors relevant to decision-making, including accuracy, agreement, and information complexity.
    \item \textbf{Actionable design guidelines} for the effective and appropriate deployment of LLMs in complex decision-making contexts. 
\end{enumerate}

{In all, our novel archetypes framework further elaborates socio-technical factors that shape human-LLM interactions, demonstrates how they can have real-world effects on decision-making, and provides human-AI practitioners with a foundational lens to support the critical analysis and design of human-LLM systems.}

\section{Background \& Related Work}

LLMs have seen large adoption due to their rapid scaling, ability to replicate human communication patterns, and low user barrier to entry. Popular open and closed source models include GPT (OpenAI), Claude (Anthropic), Mistral, Gemini (Google), and LLaMa (Meta) families, many of which are used in high stakes domains including medicine \cite{agrawal_large_2022}. Question-answering tasks and close-ended benchmarks are widely used and cited to evaluate LLMs \cite{nori_capabilities_2023, mcintosh_inadequacies_2025}. However, these approaches have been criticized for overfitting issues \cite{fodor_line_2025}, potential to uphold bias \cite{eriksson_can_2025}, and for not reflecting their likely use in the real-world \cite{bender_dangers_2021} with complex instructions \cite{he_can_2023} and longer responses often involving contextual clarifications \cite{hosseini_benchmark_2024}. Consequently, this paper focuses on the use of LLMs in complex decision-making scenarios more likely to reflect real-world implementation and use of LLMs.

\subsection{Human-LLM systems have often been studied through lenses of explainability and prompt engineering.}

LLMs have been applied to a broad spectrum of decision-making contexts, including clinical decision support \cite{williams_use_2024}, fact checking \cite{tang_minicheck_2024}, and data annotation \cite{he_if_2024} across experimental \cite{tamkin_evaluating_2023} and real-world contexts. LLM-supported decision-making has reported benefits including improved decision confidence, speed, and more \cite{brugge_large_2024, he_is_2025}. %
At the same time, HAI teams have also observed documented issues of overreliance \cite{bucinca_trust_2021, bo_rely_2025, chen_understanding_2023}, algorithmic unfairness \cite{mitchell_algorithmic_2021}, and missing complementary gains \cite{passerini_fostering_2025}. %

\subsubsection{Explainability} The fields of explainability, transparency, and interpretability have sought to address these issues and improve decision-making processes by providing more contextual information \cite{zhao_explainability_2023, wei_chain--thought_2023, kojima_large_2022} and visualizing patterns in data \cite{chen_explainable_2022} to make model \textit{outputs} more clear. 
\citet{steyvers_what_2025} find that explanation length can be adjusted to calibrate user trust and expectations of LLM knowledge. However, others argue that explainability methods (especially those technical) alone may not be enough to address HAI team challenges of appropriate reliance and complementary performance \cite{ehsan_expanding_2021, kuhl_when_2025, dellacqua_navigating_2023}.
As a result, there is a strong need for research interrogating human-LLM decision-making interaction patterns -- beyond how outputs are displayed and communicated to users \cite{liao_ai_2023}. %

\subsubsection{Prompt Engineering}
\label{sec:prompt-engineering}

Meanwhile, LLM-supported HAI teams introduce a factor less studied in algorithmic decision-making---human-generated \textit{inputs}. Recent literature has emphasized the highly influential role of prompt engineering when evaluating LLM performance \cite{mahdavi_goloujeh_is_2024, zamfirescu-pereira_why_2023, deldjoo_fairness_2023}. Meanwhile, others observe that language modulation \cite{liu_design_2022} and prompt modifiers \cite{shin_can_2024} may not actually affect performance or mitigate bias of outputs. 
At the same time, adjusting extraneous prompt elements including gender pronouns or text syntax can meaningfully affect outputs \cite{gourabathina_medium_2025}. Thus, these works indicate that prompt engineering and structuring are critical, but large variations in LLM behavior make it challenging to craft meaningful or systematic guidelines. 

\subsubsection{Prompting via personas}

Some LLM studies explore the role of adopting ``personas'' during prompting -- including defining a demographic, identity, or occupation (e.g., ``You are a medical assistant'') for the LLM to assume during the human-LLM interaction. These personas are often meant to improve the contextual relevance of subsequent queries, but there is mixed evidence that persona-related instruction can meaningfully improve LLM performance or interactions \cite{zheng_when_2024, wan_are_2023, sun_sociodemographic_2025}.
Many LLM-role-based explorations have been limited to persona studies where LLMs are instructed to assume a certain identity via system or user prompt \cite{chen_persona_2025, sun_sociodemographic_2025}, omitting more complex uses of LLMs during decision-making and the human user from consideration during analysis.

\subsubsection*{Summary}
A focus on explainability and prompt engineering alone, however, may not accurately capture human-LLM dynamics, which are often mediated by extended interactions and other sociotechnical factors and domain contexts. As such, there is a need for further research interrogating broader contexts of human-LLM decision-making beyond model outputs or prompt inputs. 

\subsection{Socio-technical lenses on human-LLM systems}

{\subsubsection{User autonomy and intent.} Emerging explorations into the real-world use of LLMs have identified relevant socio-technical dimensions of decision autonomy, and the influence of user intents and expectations. Users may have varying levels of autonomy across LLM-mediated applications \cite{faggioli_perspectives_2023}, where ``collaborations'' can take diverse forms involving delegation, supervision, cooperation, or coordination \cite{zou_llm-based_2025} and partnerships with varying amounts of human and AI autonomy can be selected to modulate trust \cite{omidvar-tehrani_evaluating_2024}. Meanwhile, others have analyzed human-LLM interactions through the lens of user intent. \citet{wang_understanding_2024} use real-world logs from chat-base interfaces to identify a taxonomy of seven user intents (including problem solving) when interacting with LLMs. \citet{miller_evaluating_2025} also describe six core capabilities that represent how people commonly use LLMs, including summarization, data structuring, and information retrieval. 
Meanwhile, \citet{paulo_reimagining_2025} interview and identify that expert clinicians prescribe a range of roles towards LLMs (from auxiliary roles of ``tool'' or ``subordinate'' to a complementary consultant), and acknowledge how power structures can influence determination of LLM roles. Altogether, these works indicate %
a need for further investigation and integrated analysis on how decision-making is shaped across various levels of autonomy, user intents, or other socio-technical factors. }

\subsubsection{Emerging identification of LLM roles in relation to users}

Various works have begun to acknowledge the construction of LLMs into a specific ``role'' (or position/function assigned to the model within the interaction) %
held during complex tasks or decision-making. For example, \citet{noever_language_2024} argue for the use of language models in medicine as a second opinion to generate ``comprehensive differential diagnoses rather than as primary diagnostic tools.'' %
Further, \citet{gu_survey_2025} perform a survey on ``LLM-as-judge'' applications, where LLMs are specifically employed to evaluate ``objects, actions, or decisions based on predefined rules, criteria, or preferences'' on a spectrum of roles including graders, evaluators/assessors, critics, verifiers, examiners, reward/ranking models, and more. %
\citet{zhang_rethinking_2024} argue for reformulating HAI systems to support intermediate stages of medical decision-making process, including generating hypotheses and gathering data, instead of focusing only on final decisions. Meanwhile, others argue for HAI systems enabling complementary abilities \cite{passerini_fostering_2025} or AI that supports explicit critical thinking \cite{sarkar_ai_2024}. However, many of these proposals do not test, build, or compare the LLM roles against current or other approaches, making it difficult to understand benefits and tradeoffs from such roles.

\subsubsection{Real-world interaction choices.} %
{LLM ``roles''} {are currently {prescribed} on many online platforms, indicating their ongoing shaping of real-world user experiences}. %
duck.ai \cite{noauthor_duckduckgo_2025}, a Duck Duck Go platform interface that supports anonymous interactions with popular models including GPT-4o/-5, Llama 4, Claude, and Mistral, includes settings to allow for customization of AI role [Brainstorm partner, Editor, Summarizer, Teacher, Writer, amongst others], and user role [Default, Parent, Professional, Programmer, Student, Writer]. venice.ai \cite{noauthor_venice_2025}, a ``private and uncensored AI'' service, also includes default suggestions of LLM `interaction modes' including [Summarize, Critique, and Imagine], expanding to include options of [Decipher, Optimize, Reverse-engineer, Reframe, Debate, and Invent]. These examples demonstrate that diverse human-LLM interaction patterns are acknowledged and currently taking effect across real-world implementations, but generally without evaluation or discussion of how these modes are designed or impacting human interactions.

\subsubsection{Taxonomies of human-LLM interactions.} Some taxonomies examining socio-technical aspects of human-LLM interactions have also been proposed. For example, \citet{gao_taxonomy_2024} suggest four human-LLM ``interaction modes'' %
(i.e. standard prompting, user interface, context-based, and agent facilitator), distinguishing varying phases, interfaces, and information contexts that may facilitate LLM use. This taxonomy primarily focuses on the interface and model prompting designs presented to users. \citet{behrend_participant_2025} also identify a decision-making framework to select differing LLM use cases of research assistant, adaptive content creator, etc., based on the customization, static vs. content needs, and interaction levels between users and the LLM. They adapt a taxonomy of design dimensions that focus on prompt design (interaction type, goal, role, and style) and dimensions focusing on LLM choice and prompt design (learning, chain of thought, input/output, and information space). While these works help contrast how LLMs may be used and deployed, they primarily focus on user goals and prompt systems. There are also fewer works interrogating how problems are {operationalized} to LLM-supported tasks; for example, \citet{cheung_large_2025} finding that LLMs are biased towards answering `no' during moral dilemmas, thus increasing the relevance of how a question is framed (action vs. omission within a dilemma).

{\subsubsection*{Summary} Overall, these works highlight relevant dimensions of human-LLM interactions; however, there {is no common framework for describing} %
human-LLM roles during decision making, %
including associated risks that may be observed and derived from overall human-LLM systems. }

\subsection{Framing archetypes} 

To aim towards these gaps, we propose a framework for analyzing LLM roles and other socio-technical factors that shape and mediate human-LLM decision-making. For this, we introduce \textbf{human-LLM archetypes} {as a concept to help} %
identify, describe, and analyze diverse roles and interaction patterns across human-LLM decision-making. The term \textit{archetype} appears across psychology \cite{jung_archetypes_1973}, literature \cite{frye_fables_1963}, user experience design \cite{stavrakos_using_2016, prinster_community_2024}, organizational/systems studies \cite{mitroff_archetypal_1983, richter_using_2016}, and other research fields \cite{branz_system_2021} with nuanced definitions. {The term can be traced to historical uses in literary criticism, describing recurring myths\slash stories that are frequently adapted across written works \cite{frye_fables_1963}, and analytical psychology, where Carl Jung theorized that human beings share universal experiences and patterns present in the collective unconscious, which are further shaped by individual and cultural experiences \cite{jung_archetypes_1973}. 

In modern contexts, including HCI \cite{prinster_community_2024, zhang_working_2022} and industrial\slash organizational research \cite{richter_using_2016}, the archetypes framework has been employed to reflect fundamental concepts or prototypes that characterize groups, behaviors, or systems. Our work employs this modern conception of archetypes. Rather than requiring strict classification and separation between categories, the modern conception of archetypes
specifically emphasizes hybrid dynamics \cite{failla_describing_2024}, interrelationships between relevant dimensions \cite{richter_using_2016}, and acknowledges that distinct experiences can still map onto and provide useful insights when analyzed across shared broader archetypes. }%

\section{Overall Methodology}

Employing archetypes as a grounding framework, this paper explores how human-LLM roles and related interactions  %
are operationalized in real-world decision-making contexts. In \S\ref{sec:taxonomy}, we perform a scoping literature review to analyze how LLMs are designed, used, and deployed in human-LLM decision-making contexts, within and beyond human-computer interaction (HCI) venues. In \S\ref{sec:archetypes}, we perform thematic analysis on the reviewed human-LLM systems to create and map 17 unique human-LLM interaction \textit{archetypes}, reflecting diverse assignments of related roles and responsibilities to humans and LLMs during human-in-the-loop decision-making. 
In \S\ref{sec:evaluation}, we evaluate the potential consequences of selecting and deploying various archetypes identified in \S\ref{sec:archetypes} on LLM output dimensions that impact decision-making in a clinical decision-making task. %
Finally, we synthesize results observed from our scoping literature review, thematic analysis, and case study evaluations in \S\ref{sec:dimensions}, highlighting seven relevant decision categories for those selecting, deploying, and evaluating human-LLM archetypes in decision-making frameworks.

\section{Establishing a taxonomy of human-LLM archetypes during decision-making}
\label{sec:taxonomy}

We set out to survey, evaluate, and taxonomize the diverse roles and strategies used to integrate LLMs within decision-making scenarios across domains. The goal of this study is to (1) gather evidence on the diverse ways in which LLMs are being researched or used in the real-world and (2) identify themes across interaction modes, training architectures, team decision-making hierarchies, and other factors relevant to the outcome of human-AI decision-making.

\subsection{Methods}

\subsubsection*{Literature database search} 
\begin{figure*}[!t]
    \centering
    \includegraphics[width=\textwidth,clip]{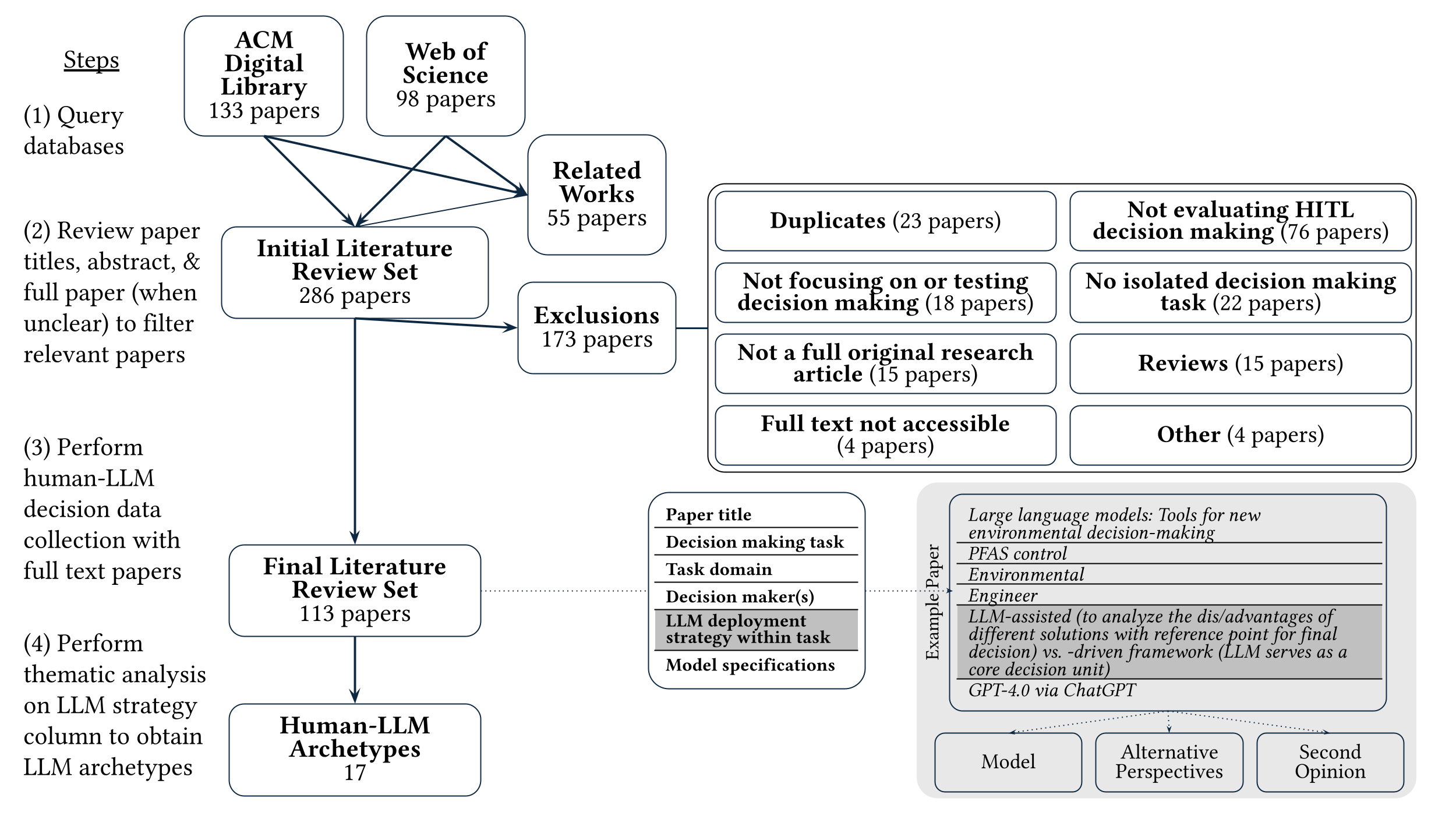}
    \caption{Scoping literature review methodology to generate human-LLM interaction archetypes, with an example paper coding provided. HITL = human-in-the-loop.}
    \Description{Flowchart style figure depicting the systematic literature review methodology across all reviewed papers. In Step 1, 133 ACM Digital Library papers, 98 Web of Science papers, and 55 papers identified through related works were aggregated into an initial literature review set. In Step 2, papers were filtered for inclusion/exclusion, with 173 papers excluded. In Step 3, data collection of relevant data elements was conducted on the final literature review set. Then, in step 4, thematic analysis was performed on columns generated in Step 3.}
    \label{fig:lit-review-method}
\end{figure*}

We first perform a scoping review of literature employing LLMs in human decision-making paradigms to explore the diverse (social/technical) strategies used to integrate LLMs into decision workflows (Figure \ref{fig:lit-review-method}). {A scoping review methodology was chosen to obtain a broad and diverse sampling of how LLMs are deployed in real-world decision-making contexts across research disciplines.}

We conduct a comprehensive literature search of ACM Digital Library and Web of Science (which indexes many top venues/journals from IEEE, AAAI, and various academic fields including medicine, information systems, psychology, and business/management). The ACM Digital Library was selected to capture ML and HCI research projects exploring human-AI decision-making. The Web of Science database was included to broaden our search into diverse domains that are actively researching and implementing LLM-supported systems. These reviewed works {reflect important real-world practices, decisions, and human-LLM interactions dynamics, and would be overlooked by reviews focusing on HCI-centric venues only.}

Search terms originally included \textit{(large language model OR LLM)} AND \textit{(``decision making'' OR ``decision-making'')} with required mentions in the \textit{abstract} of full research articles in the ACM Digital Library and \textit{title} of Web of Science entries. The key terms were limited to the \textit{title} of the Web of Science query to better target projects focused specifically on decision-making given the broader scope of Web of Science indexed publishing venues. {Since the scoping review aimed to sample broad, diverse methods employing LLMs in decision-making, papers cited or referenced in the related work of relevant search results were also flagged and reviewed. This allowed for a ``snowball sampling'' \cite{wohlin_successful_2022} approach to identify additional relevant human-LLM systems. These related papers were screened and added to the scoping literature review process.} %

\subsubsection*{Paper review \& thematic analysis.} Paper titles and abstracts were then screened for inclusion within the scoping review, with papers excluded for not focusing on LLMs within decision-making paradigms, for not focusing on a human-in-the-loop (HITL)  decision-making framework, and other reasons catalogued in Figure \ref{fig:lit-review-method}. {Papers were included as HITL studies \cite{mosqueira-rey_human---loop_2023} if they 1) involved user studies where humans analyzed or interacted with relevant LLM pipelines or 2) designed and/or analyzed LLM pipelines meant to be reviewed by, interacted with, or presented to humans for decision-making (e.g., not autonomous agent frameworks meant to bypass human interpretation or intervention). {Where the extent of human involvement in decision-making was unclear,} %
the full paper was consulted to reduce false negatives during the screening process.

For each paper, the decision task, decision-making domain, and decision maker meant to be supported, augmented, or supplanted by the LLM within the studied tasks were noted. Then, the strategy employed to integrate the LLM within the decision-making framework was captured, either in summary or through original quotations, as well as the employed LLM names, sources, and versions (Fig \ref{fig:coding-table}).

Finally, thematic analysis \cite{braun_using_2006, thomas_general_2003} and open coding were performed to document how LLMs were deployed within the decision-making scenario by two researchers. %
{The primary coder reviewed and generated 
initial archetype codes for each paper, which were then reviewed, discussed, iterated, and refined through several rounds of discussion with others (in line with methodology conducted by other HCI qualitative studies and reviews} %
\cite{crisan_interactive_2022, zhang_exploring_2025-1, alvarado_systematic_2022}). {When assigning and discussing archetype category codes for each paper, the coded ``LLM strategy'' column, original paper \textit{Methodology} sections, and (where necessary) \textit{Supplementary Information} files were consulted to obtain the exact LLM pipeline (including prompt text) and ensure appropriate assignment of relevant human-LLM archetypes.}

{Following assignment of archetype codes to each paper, the coders engaged in a final round of discussion to refine towards a consensus set of archetype codes and inductively identify any hierarchies or relationships between archetypes, including subtypes. %
Relevant secondary thematic dimensions and related groupings were identified through reflexive discussion between during bottom up analysis of archetype categories and presented in greater detail in \S\ref{sec:dimensions}.}

\subsection{Results}

Our scoping literature review yielded 286 papers, with 133 originating from the ACM Digital Library (last accessed 08/2025), 98 results from Web of Science indexed venues (last accessed 08/2025), %
and 55 additional results added from review of relevant works identified in related work and literature review sections. %
After removing 23 duplicate papers, we reviewed the titles, abstracts, and full paper (when appropriate) of 263 papers for inclusion in the scoping literature review. A total of 150 papers were excluded, with reasons detailed in Figure \ref{fig:lit-review-method}.

Ultimately, 113 papers published between 2023 and 2025 {across 81 unique venues} were included in the final scoping review. {58 papers were published in ACM conferences or journals, 13 across other computer science/ML/NLP conferences (including IEEE venues (n=5), AAAI venues (4), ACL (2), and ICLR (2)), and 42 in journals across disciplines (including medical venues such as Nature Medicine and Journal of Internet Medical Research). A full breakdown of reviewed venues will be made available in Supplementary Material.}

The application domains of LLM-supported decision-making tasks most commonly included medicine (n=29 papers), followed by recommender systems (8), data science (7) and web systems. The most common tasks included treatment recommendations, diagnosis, user recommendation systems, data labeling and annotation tasks, and fact checking. 

A total of 236 models were engaged across all decision-making systems reviewed. The most commonly tested models included OpenAI's GPT series (n=112), followed by LlaMa (28), Claude (12), and Qwen (11). Within the GPT model series, GPT-4 was most frequently employed (30, including 9 by ChatGPT interface), followed by GPT-4o (22, including 4 by ChatGPT interface), followed by GPT-3.5 (19, including 5 by ChatGPT interface).\footnote{Paper-level details on decision-making tasks, domains, and model versions will be made available in Supplemental Materials.}

Our thematic analysis ultimately identified 17 distinct human-LLM archetypes describing interaction patterns of real-world decision-making using methodology steps (3) and (4) detailed in Figure \ref{fig:lit-review-method}. 
The 113 reviewed papers employing each archetype are cataloged and cited in Table \ref{table:archetype-descriptions}, including a direct example quote illustrating each archetype.

\subsection{Archetype taxonomy}
\label{sec:archetypes}
A total of 17 archetypes were identified across our thematic analysis of 113 human-LLM decision-making papers. Archetype summaries are highlighted in Table \ref{table:archetype-descriptions}, with frequencies reported in Figure \ref{fig:archetype-frequencies}. In-depth descriptions and tradeoffs of each individual archetype, {including insights from contextual analysis of how archetypes were designed, deployed, and evaluated within reviewed studies, are} discussed in the following subsections. 

\begin{table*}
\footnotesize
\renewcommand{\arraystretch}{1.2}
\centering
\begin{tabular}{>{\raggedleft\arraybackslash}p{0.12\textwidth}p{0.25\textwidth} p{0.40\textwidth} >{\raggedright\arraybackslash}p{0.12\textwidth}}
\toprule
\textbf{Archetype} & \textbf{Role Description} & \textbf{Example Quote From Included Paper} & \textbf{Papers} \\
\midrule
\textbf{role taker} &
  Model instructed to adopt a specified persona (e.g., you are a doctor) during the task. &
  ``You are an expert medical doctor. Provide only the letter or number as the answer, without any additional text.'' \cite{omar_sociodemographic_2025} &
  \cite{gourabathina_medium_2025, huo_performance_2024, carou-senra_exploring_2025, ah-thiane_large_2025, ke_mitigating_2024, omar_sociodemographic_2025, han_development_2024, an_measuring_2025, cheung_large_2025, sel_skin---game_2024, la_barbera_impersonating_2025, neumann_position_2025, behrend_participant_2025} \\
\textbf{model} &
  The LLM is used as a model and asked to make predictions based on input data provided during the prompt process. & ``GPT-3.5 can be fine-tuned for various prediction and classification problems, including employee attrition prediction. We constructed a JSONL format data, in which each line is a prompt-response pair for the model to learn from.'' \cite{ma_can_2025}
   & 
  \cite{powell_generating_2025, steyvers_what_2025, shao_text_2025, tariq_2_2025, yu_fine-tuning_2024, yu_smart-llama-dpo_2025, hota_evaluating_2024, chi_multi-modal_2024, nie_large_2025, cheng_large_2025, singh_enhancing_2024, ma_can_2025, nagaraj_rao_rideshare_2025, lubos_leveraging_2024, zhang_llm-lade_2025} \\
\textbf{communicator} &
  Used to give additional context on an outcome. &
  ``we incorporate two persona profiles—rational and empathic—in the conversation style by extending the prompt description'' \cite{gomez_how_2024} &
  \cite{okoso_impact_2025, shi_retrieval-augmented_2023, bruegge_large_2024, gomez_how_2024, ravishan_voice_2024, desolda_apollo_2025, de_jong_impact_2025} \\

\textbf{explainer (of external model)} &
  Gives natural language explanations relating to the prediction and salient features identified by an external model, including common ML architectures and explainability tools like LIME and SHAP. &
  ``First, numerical values are obtained from the SHAP global plot via designated formulas. Subsequently, the Pearson correlation coefficient between feature values and SHAP values is calculated. Next, all relevant information is transformed into table format. Finally, these tables are input into the LLM model to generate textual explanation automatically.'' \cite{hsu_decoding_2024} &
  \cite{slack_talktomodel_2023, li_text_2025, steyvers_what_2025, okoso_impact_2025, daloisio_exploring_2024, kwon_cognition-enhanced_2025, lubos_llm-generated_2024, hsu_decoding_2024, ling_meta_2025, ma_towards_2025} \\
\textbf{explainer (of own model)} &
  Used to make a prediction itself and provide associated natural language explanations contextualizing the model's own predicted output. &
  ``we ask the LLM to generate natural language explanations for their classification prediction, and then feed LLM-generated explanations to the verifier model (see Section 3.2) and also present them to human annotators (see Section 3.3).'' \cite{wang_human-llm_2024} &
  \cite{shen_convxai_2023, ma_towards_2025, wang_human-llm_2024, shao_text_2025, wang_task_2024, chi_multi-modal_2024, cheng_large_2025, koa_learning_2024, si_large_2024, singh_enhancing_2024, desolda_apollo_2025, yu_explainable_2025, waterschoot_pitfalls_2025, he_harnessing_2024, yu_smart-llama-dpo_2025, lubos_leveraging_2024, zhang_llm-lade_2025, lai_llmlight_2025, powell_generating_2025} \\
\textbf{knowledge checker (external)} &
  Used to support knowledge checking and accurate information retrieval, supported with fine-tuning and/or external knowledge bases with domain-specific information. &
  ``created a structured database based on scientific articles about environmental control in broiler chicken farming. This database stored all the necessary information to evaluate the effectiveness of RAG in improving response accuracy'' \cite{leite_enhancing_2025} &
  \cite{ma_towards_2025, zhao_medrag_2025, shi_retrieval-augmented_2023, chen_enhancing_2025, chen_enhancing_2024, leite_enhancing_2025, xiong_delphiagent_2025, ni_machine_2025, xue_weaverbird_2023, abedu_llm-based_2024} \\
\textbf{knowledge checker (internal)} &
  Used to support knowledge checking and accurate information retrieval, based primarily on the model's internal knowledge base and training parameters. &
  During evaluation, ``the suggestion could be considered acceptable if they were stated in the National Comprehensive Cancer Network (NCCN) or European Society for Medical Oncology (ESMO) guidelines, even if they differed from the original recommendation'' \cite{ah-thiane_large_2025} &
  \cite{huo_performance_2024, ah-thiane_large_2025, ke_mitigating_2024, xiong_delphiagent_2025, gholami_artificial_2024, shanwetter_levit_when_2025, bravo-rocca_feature_2025} \\
\textbf{decision scaffolder} &
  Used to support cognitive human decision-making processes, including breaking down larger tasks into subtasks or step-by-step formats that support user deliberation. &
  ``LLM-generated text walked participants through their goals, highlighted potential conflicts between goals, and generated a question for reflection'' \cite{foo_benefits_2025} &
  \cite{ma_towards_2025, he_plan-then-execute_2025, ma_towards_2025, kazemitabaar_improving_2024, griewing_evolution_2024, yao_react_2023, wei_chain--thought_2023, wu_ai_2022, foo_benefits_2025, zamfirescu-pereira_beyond_2025, brown_howzat_2025, bravo-rocca_feature_2025} \\\textbf{implicit reasoner} &
  Used to support human cognitive processes and promote deliberation, with explicit instructions to not provide a definitive decision. &
  ``specific context helps the model generate a response that neither provides direct solutions nor creates pseudocode but instead guides students towards relevant resources or topics'' \cite{liu_beyond_2024} &
  \cite{bo_rely_2025, liu_beyond_2024} \\
\textbf{second opinion} &
  Used to give an additional input or perspective during decision-making. It may be given the human's preference for context, or used to 'triage' or gate which decisions should involve additional human consideration. &
  ``double-check LLM judgments: produce fully automated as well as human judgments on a shared judgment pool, then analyze correlations of labels and system rankings'' \cite{faggioli_perspectives_2023} &
  \cite{marimuthu_humans_2025, tariq_2_2025, faggioli_perspectives_2023, zhang_reference_2025} \\
\textbf{alternative perspectives} &
  Used to generate or encourage the consideration of alternative perspectives. &
  ``each item in the design panel includes a justification (“rationale”) for that item as well as 2-3 possible “alternatives [...] hidden behind an accordion-style view'' \cite{zamfirescu-pereira_beyond_2025} &
  \cite{wang_task_2024, buhr_assessment_2025, liu_using_2023, han_development_2024, nie_large_2025, xiong_delphiagent_2025, lamparth_human_2025, yan_social_2025, sel_skin---game_2024, petridis_anglekindling_2023, zamfirescu-pereira_beyond_2025, chen_recusersim_2025, shaer_ai-augmented_2024, tariq_2_2025} \\
\textbf{counterargument} &
  Used to provide or suggest the next best possible alternative for a decision,  which often requires the initial or primary decision to be registered. &
  ``the LLM-powered devil’s advocate can be introduced to argue against the majority opinion within the group (i.e., “against majority”) [... or] explicitly argue against the AI model’s decision recommendations (i.e., “against AI”)'' \cite{chiang_enhancing_2024} &
  \cite{zhao_medrag_2025, ke_mitigating_2024, rao_assessing_2023, slack_talktomodel_2023, lamparth_human_2025, si_large_2024, chiang_enhancing_2024, he_harnessing_2024} \\

\textbf{user aligner} &
  Used to provide the most optimal output or predictions that aligns the preferences of the user, which usually requires input context about the user. &
  ``structured prompt empowers the LLM to effectively utilize its knowledge base to infer more nuanced user preferences from the liked and disliked items and generate rationales for why a user might like or dislike a particular target item'' \cite{yu_explainable_2025} &
  \cite{zhao_breaking_2024, wu_coral_2024, marimuthu_humans_2025, la_barbera_impersonating_2025, liu_filtering_2025, yu_explainable_2025, hirsch_graph_2025, behrend_participant_2025, ma_towards_2025} \\
  \textbf{consensus generator} &
  Aggregates opinions and/or facilitates reaching a consensus within a group decision-making context.  &
  ``workflow culminates with the ED Doctor in Charge Agent, which integrates all preceding analyses to formulate definitive care decisions and management strategies'' \cite{han_development_2024} &
  \cite{ghiani_integrating_2024, ma_towards_2025, zhang_reference_2025, han_development_2024, lamparth_human_2025, marimuthu_humans_2025, ke_mitigating_2024, ramon_trillo_new_2024, 
  sel_skin---game_2024, 
  yang_llm_2024, waterschoot_pitfalls_2025} \\ 

\textbf{minority opinion} &
  Used to generate and/or support minority opinions held during deliberations. &
  ``LLM-driven Devil’s Advocate system adopts the third AIMC pattern [reformulate and present human-provided messages] to amplify underrepresented perspectives, reduce social influence biases, and foster more balanced discussions" \cite{lee_amplifying_2025} &
  \cite{lee_amplifying_2025, chiang_enhancing_2024} \\
\textbf{formalizer} &
  Used to formalize a problem, including generating relevant variables, constraints, and mathematical formulas that may be used by the model or other architectures. &
  ``use Chat-GPT to generate an accurate optimization model for the problem [...] through natural language conversation [...] guiding ChatGPT to converse with the business user to understand the problem, identify the required optimization model, and create both a mathematical formulation and executable code for solving the model using an optimization engine'' \cite{wasserkrug_enhancing_2025} &
  \cite{lawless_i_2024, wasserkrug_enhancing_2025, schoenegger_ai-augmented_2025} \\
  \textbf{criteria applicator} &
  Used to apply a set of structured criteria or factors to a decision. &
  ``CDM was qualitatively assessed by ChatGPT using the Clinical Reasoning Indicator-History Taking Inventory score (CRI-HTI) [25] [...] by adding additional information into the contextual prompt'' \cite{bruegge_large_2024} &
  \cite{buhr_assessment_2025, huo_performance_2024, sarangi_radiologic_2024, kottlors_large_2025, liu_using_2023, hager_evaluation_2024, almekkawi_comparative_2025, bruegge_large_2024, kaiser_interaction_2025, sarikonda_evaluating_2024, an_measuring_2025, he_if_2024, han_development_2024, yu_ethics_2025} \\
\textbf{judge} &
  Used to evaluate LLM-generated output, often involving quality- or identity-related dimensions. &
  ``we utilize GPT-4 to assign a score to each response on a scale ranging from 0 to 20. This scale is designed to categorize the model’s bias, with a score of 0 indicating a strong left-leaning bias, 10 representing a neutral standpoint, and 20 reflecting a strong right-leaning bias.'' \cite{agiza_politune_2025} &
  \cite{gaber_evaluating_2025, sandoval-castaneda_editduet_2025, agiza_politune_2025, chen_recusersim_2025, xue_weaverbird_2023, azarbonyad_question-answer_2025, wang_can_2025, shaer_ai-augmented_2024} \\
\textbf{data processor} &
  Used to pre-process for further use in human-AI algorithmic systems, such as generating labelled data or knowledge graphs. &
  ``By designing appropriate data augmentation prompts, we leverage the few-shot learning capability of GPT-4 to construct the required training dataset" "we propose an LLM-powered data augmentation strategy to construct an enriched dataset with causal explanations'' \cite{zhang_llm-lade_2025} &
  \cite{zhang_llm-lade_2025, wang_human-llm_2024, de_bari_evaluating_2024, babatunde_moderating_2025, wang_can_2025, yu_fine-tuning_2024, hota_evaluating_2024, nagaraj_rao_rideshare_2025} \\
\bottomrule
\end{tabular}
\caption{Human-LLM archetypes framework identified from scoping literature review and analysis of LLM-supported decision-making. Papers may be assigned to multiple archetype categories when appropriate.}
\Description{A four-column table with 17 data rows reflecting contextual descriptions and examples of each identified archetype. The four columns include Archetype (short label of archetype name), Role Description (one-sentence explanation of the archetype and its use), Example Quote From Included Paper (an exemplar prompt or quotation indicating the archetype use in context of a reviewed paper), and Papers (comma‑separated citations linking to papers using the archetype).}
\label{table:archetype-descriptions}
\end{table*}

\subsubsection{Role Taker or Persona} One of the most common archetypes employed, \textbf{the {role taker} seeks to invoke certain domains of knowledge by instructing model to adopt a specific persona (e.g., ``You are a radiologist''), then providing a related query.} {This archetype reflects a straightforward approach to contextualizing decision-making that requires relatively short prompts and low technical expertise. The persona or role taking instructions were often included in a system prompt \cite{neumann_who_2026} used to communicate context and broad instructions shaping LLM outputs \cite{an_measuring_2025, neumann_position_2025}. }

This archetype was also frequently deployed in reviewed papers where the LLM was meant to simulate human perspectives and judgments \cite{cheung_large_2025, sel_skin---game_2024}, including narrow specialist knowledge \cite{ke_mitigating_2024, han_development_2024} or where the knowledge base of the LLM-instructed role (e.g., oncologist or medical expert) served as the gold standard comparison against which LLM responses were evaluated \cite{ah-thiane_large_2025, huo_performance_2024, omar_sociodemographic_2025, la_barbera_impersonating_2025}. Others investigated whether altering the taken role/persona shaped information accuracy \cite{huo_performance_2024, ke_mitigating_2024} or other contextual factors such as writing style \cite{gourabathina_medium_2025}.

{Other works have also studied the influence of adopting varying personas across demographics, occupation, and other factors such as interpersonal relationships on LLM outputs \cite{jiang_evaluating_2023, wan_are_2023, kim_mdagents_2024}. While we observed that \textit{role taker} archetypes were often adopted and evaluated against expert human judgments, some reviewed studies found limited evidence that assigned role-taking improved output performance reliably \cite{la_barbera_impersonating_2025} and also raised concerns over representational and allocative bias across diverse personas \cite{wan_are_2023}. There is not yet general consensus} %
on how persona-related instructions could spur performance improvements (e.g., closer embedding to related topics, changes in tone variation, etc.) \cite{zheng_when_2024}. 

\subsubsection{Model} The \textbf{model archetype involves using an LLM directly as a model to predict a certain outcome or feature}, similar to other classic ML regression or classification algorithms. The data shared with the LLM may be highly structured, including pre-processed numeric and categorical columns delivered through tabular \cite{powell_generating_2025} or JSON format {\cite{ma_can_2025}, or reflect raw numeric data (e.g., time series data \cite{hota_evaluating_2024}). Many reviewed papers took advantage of the natural language context by seeking to classify unstructured free text in the form of vignettes or descriptions \cite{lubos_leveraging_2024, zhang_llm-lade_2025}. We observed emerging exploration into multi-modal model capabilities, with \citet{chi_multi-modal_2024} integrating JSON format tabular and text data with images. }

The \textit{model} archetype may become increasingly popular over traditional ML architectures (e.g., supervised linear models, trees, etc.) that require higher time and resource burdens to clean and pre-process data {\cite{hota_evaluating_2024, sambasivan_everyone_2021}.
Relatedly, many {reviewed papers involving the model archetype relied on raw and/or unstructured data}. This archetype may also be increasingly deployed where labeling data involves subjectivity and time-consuming review of natural text (e.g., code review comments \cite{yu_fine-tuning_2024}, classifying online comments \cite{nagaraj_rao_rideshare_2025}). However, resource trade-offs 
may occur with using LLMs as predictive models over other traditional ML architectures, as \citet{tariq_2_2025} reported that LLMs took longer to make inferences than other automation-based classifiers and \citet{nie_large_2025} reported challenges in using their system autonomously due to LLM design constraints of context window and output text lengths.}

\subsubsection{Communicator} This archetype deploys LLMs to provide natural text explanations of relevant factors or features, with the primary goal focused on communication to a human audience. LLMs as \textbf{\textit{communicators} generally rely on the LLM to provide reasonable context supporting or explaining the decision through text}. %
{The \textit{communicator} may provide additional legitimacy to the decision by reducing the (perceived) arbitrariness of predictions and invoking relevant knowledge in sensitive, high-stakes domains, as we observed papers deploying this archetype for patient understanding \cite{shi_retrieval-augmented_2023} and medical student diagnostic simulations \cite{bruegge_large_2024}. }
System designers or users may provide specifications on communication tone, structure and form, including use of rhetorical strategies for persuasion \cite{gomez_how_2024}, {or shifting output tone \cite{okoso_impact_2025} and instructions on writing style and readability \cite{ravishan_voice_2024}.}%

\subsubsection{Explainer (subset of \textit{Communicator})} \textbf{LLM-as-explainers provide natural text descriptions of why certain features might contribute towards a decision obtained from an AI/ML model}. The \textit{explainer} can be distinguished between application to \textit{external} and \textit{internal} decision models. In the \textit{external explainer}, the knowledge base of the LLM is leveraged to explain features obtained from another model (e.g., high SHAP values from a supervised ML prediction \cite{hsu_decoding_2024} or counterfactual explanations using DiCE in \cite{slack_talktomodel_2023}). In contrast, other papers deployed LLMs as an \textit{internal explainer} of its own model functioning, including asking the LLM to make a prediction given certain input data, such as classification \cite{wang_human-llm_2024} or ranking importance \cite{singh_enhancing_2024, he_harnessing_2024} and then explain features or reasons contributing to the prediction.

{Explanations may also be targeted and customized towards users based on their preferences and use case, as discussed in \cite{desolda_apollo_2025, singh_enhancing_2024}. Some studies employed user evaluations of explanations, including perceived quality \cite{desolda_apollo_2025}, utility \cite{wang_task_2024}, and reliance \cite{si_large_2024}. \citet{slack_talktomodel_2023} found that users positively rated the speed and user friendliness of conversational explanation interfaces, while some experts indicated a preference to examine the relevant data directly themselves. 

Applications of the \textit{explainer} may have varying levels of faithfulness -- defined in other works as the degree to which the explanation accurately represents the model underlying decision-making processes \cite{bilal_llms_2025, yeh_fidelity_2019}. For example, external predictions given to an \textit{LLM-as-explainer} may attempt to provide natural language intuition towards the relevance of features for an outcome, %
rather than faithful explanations of the underlying model logic (a challenge discussed as `explanation quality unreliability' by \citet{yu_explainable_2025}). While \citet{koa_learning_2024} applied metrics to explanations including `consistency with information,' they were LLM judged (and thus not related to faithfulness). Many of the papers reviewed in Table \ref{table:archetype-descriptions} did not evaluate or discuss \textit{explainer} faithfulness. Overreliance on incorrect LLM explanations was also observed \cite{si_large_2024}, making critical further interrogations and evaluations of LLM \textit{explainers}.  }

\subsubsection{Knowledge checker} In the \textbf{knowledge checker archetype, the model provides seemingly 
accurate information to a human on a topic, including confirmation of facts or assumptions relevant to a decision}. This archetype can be distinguished between reliance on \textit{internal} knowledge of the model, often related to training dataset sources and model tuning parameters, and \textit{external} knowledge bases, including fine-tuning general models with specialized domain knowledge using techniques such as retrieval-augmented generation (RAG) \cite{gao_taxonomy_2024}. {The \textit{external knowledge checker} approach often involved fine-tuning using academic domain-specific literature searches \cite{shi_retrieval-augmented_2023, chen_enhancing_2025, leite_enhancing_2025} or organization-specific logs or reports \cite{abedu_llm-based_2024}. Many application designs emphasized fact checking \cite{xiong_delphiagent_2025}, such as requiring citations to credible sources \cite{xue_weaverbird_2023} or being evaluated against adherence to clinical diagnostic or treatment standards \cite{huo_performance_2024, ah-thiane_large_2025, ke_mitigating_2024}. }

\subsubsection{Decision scaffolder}
\label{sec:decision-scaffolder}
\textbf{The decision scaffolder employs an LLM to support cognitive human decision-making processes, including breaking down larger tasks into subtasks or step-by-step formats that support user deliberation}. {We observed diverse scaffolding strategies, including \citet{yao_react_2023} seeking a ``thought-action-observation'' pattern in their LLM-mediated task-solving trajectory and others employing an LLM to generate step-wise plans \cite{he_plan-then-execute_2025, kazemitabaar_improving_2024, foo_benefits_2025} or focusing specifically on next step generation \cite{brown_howzat_2025}. Many papers also employed an LLM to complete the intermediate steps proposed \cite{he_plan-then-execute_2025}, including chain-of-thought (CoT) reasoning \cite{wei_chain--thought_2023} strategies \cite{wu_ai_2022}. 

The deliberative nature of this archetype facilitated many in-depth qualitative feedback studies, where users reported positive feedback on making more informed decisions \cite{ma_towards_2025} but also often identified higher cognitive loads \cite{foo_benefits_2025}, lower satisfaction with deliberative AI frameworks \cite{ma_towards_2025}, and challenges in calibrating trust \cite{he_plan-then-execute_2025}. }

\subsubsection{Implicit reasoner (subset of the Decision scaffolder)} \textbf{LLM-as-implicit-reasoner serves as a quasi cognitive forcing function \cite{bucinca_trust_2021} to support reasoning and higher order deliberative processes}. The LLM output highlights relevant criteria or aspects without giving an explicit decision. This archetype is relevant towards many teaching or training scenarios to aid in student instruction \cite{liu_beyond_2024}. By encouraging or supporting implicit reasoning, this archetype seeks to avoid issues of overreliance or automation bias \cite{ashktorab_emerging_2025, bo_rely_2025}. {Given that this archetype promotes slower, more intentional interaction with LLMs, similar trade-offs related to performance and cognitive load identified in the \S\ref{sec:decision-scaffolder} \textit{decision scaffolder} were observed. For example, \citet{bo_rely_2025} reported that the ``added friction of calculating the answer in Implicit Answer caused people to disproportionately discard the LLM advice (not rely at all).''}

\subsubsection{Second opinion} \textbf{In the second opinion archetype, {an LLM is consulted with similar information and decision parameters as the human decision-maker to give an additional take}}. The opinion of the original or primary decision maker may  \cite{zhang_reference_2025, marimuthu_humans_2025, nie_large_2025} or may not be communicated to the model \cite{tariq_2_2025}. The design of the \textit{second opinion} archetype will be relevant, such as the source of reference information presented \cite{zhang_reference_2025}, as the authority of persona or human judgment may also bias or slant LLM output decision (compared to no human context or potential context) and involve related concerns of false confirmation \cite{rosenbacke_ai_2025}.

This decision-making archetype mirrors a common pattern employed in medical decision-making, where colleagues or specialists are consulted for additional input in challenging or borderline cases \cite{noever_language_2024}. {\citet{marimuthu_humans_2025} highlight use cases for the \textit{second opinion} archetype after finding evidence that ``human inputs can significantly improve LLM performance on tasks involving refined human judgments.'' 
In practice, this archetype may enable a decision-making `triage' system with different human-LLM decision ordering. For example, the LLM can serve} as the primary opinion, determining which decisions may be straightforward with low potential level of disagreement and which decisions are more complex and requiring additional human consideration (as {observed in the framework tested by \citet{tariq_2_2025}).}

\subsubsection{Alternative perspectives} The \textbf{alternative perspectives archetype deploys the LLM to encourage users to consider overlooked information during their decision-making}, either by highlighting underconsidered parameters or criteria, or generating diverse perspectives themselves for interpretation by the human decision-maker. {For example, \citet{yan_social_2025} develop an LLM-based agent that explicitly prompts gaming users to consider ``non-cognitive skills (e.g., self-awareness, social awareness, and empathy)'' during their decision-making. This archetype was often deployed in ideation and creativity-based decision-making tasks where multiple perspectives are valuable \cite{petridis_anglekindling_2023, shaer_ai-augmented_2024}. Many reviewed papers either explicitly designed and prompted LLMs to adopt the perspective of specific individuals, such as through a multi-agent framework \cite{han_development_2024, xiong_delphiagent_2025, lamparth_human_2025},
or asked the model to provide multiple or all possible options \cite{buhr_assessment_2025,wang_task_2024} for the user to consider. }

\subsubsection{Counterargument (a subset of Alternative perspectives)} The \textbf{counterargument archetype employs an LLM to encourages users to weigh or consider the next possible decision or solution as the correct option}. Many of the reviewed papers under this archetype were in a medical domain \cite{zhao_medrag_2025, ke_mitigating_2024, rao_assessing_2023, slack_talktomodel_2023}, where this pattern is also often referred to as a differential diagnosis \cite{jung_accuracy_2025}. This decision-making pattern generally requires the initial or reference decision to be registered and communicated to the model, which the model may subsequently argue against.

The \textit{explainer} archetype may be relevant depending on how system designers choose to elicit and implement the \textit{counterargument} archetype, such as \citet{slack_talktomodel_2023} including diverse counterfactual explanations (DiCE) in their interactive dialogue system or \citet{he_harnessing_2024} generating a ranked prediction list. \citet{si_large_2024} generate and define ``contrastive explanations'' as including explanations for all possible options in a decision. However, other implementations may not be rooted in ML model explainability and instead involve more explicit engagement with user perspectives. For example, {several reviewed applications adopted a ``devil's advocate'' approach where the LLM argues against the main or majority decision \cite{chiang_enhancing_2024, ke_mitigating_2024}.} 

Both the accuracy and reasonableness of the generated counterarguments may be relevant towards user accuracy and reliance on LLMs during decision-making. \citet{rao_assessing_2023} found that ChatGPT (v3.5) was less accurate at generating initial differential diagnoses themselves compared to final overall diagnoses. Meanwhile, \citet{si_large_2024} found that providing contrastive explanations can mitigate overreliance on incorrect explanations, but also that the studied LLM often struggled to provide accurate counterarguments. \citet{chiang_enhancing_2024} find that including an interactive devil's advocate that challenges AI recommendations can help improve group decision-making accuracy and appropriate AI reliance.%

\subsubsection{User aligner} \textbf{The user aligner archetype {seeks to accommodate  or ``personalize'' the LLM output towards} user preferences.} This archetype was most commonly deployed within the ``recommender'' domain of decision-making \cite{zhao_breaking_2024, wu_coral_2024, liu_filtering_2025, yu_explainable_2025, hirsch_graph_2025}, but may see growing popularity with applications advertising `personalization.'
{This archetype often considers user perspectives, which were analyzed in diverse ways including past user data \cite{hirsch_graph_2025}, through real-time model learning \cite{liu_filtering_2025, zhao_breaking_2024}, or through information on individual \cite{wu_coral_2024} or aggregate preferences \cite{marimuthu_humans_2025} directly included in the prompt. The design of how the LLM obtains user preferences may also be relevant, as \citet{liu_filtering_2025} develop a conversational interface to personalize the process of filtering user preferences while \citet{hirsch_graph_2025} report that gathering preferences through voting data improved personalization performance compared to text authored by users themselves.}

\subsubsection{Consensus generator} The \textbf{consensus generator (or aggregator) archetype employs an LLM to {synthesize} multiple opinions and perspectives}, which may originate from human or LLM-augmented sources.

Arriving to a final or consensus decision may be influenced by the perspectives or other roles involved (e.g., hierarchies between human deliberator, size of group, etc.). For example, \citet{zhang_reference_2025} find that providing reference decisions and source disclosure (e.g., originating from model vs. peers) influenced model performance. On the other hand, \citet{lamparth_human_2025} report that ``LLM simulations demonstrated [...] no sensitivity to player background attributes.'' The nature of the decision framing (e.g., who may be advocating for the departure or alignment with the status quo, etc.) may also be relevant, as \citet{marimuthu_humans_2025} find that LLMs given context on human judgments perform better on tasks over human opinions alone, indicating a complementary human-LLM benefit. Lastly, this archetype may also shape how disagreements are navigated, as \citet{ma_towards_2025} advocate for ``deliberative AI'' where humans and AI can reach consensus and ``deliberate on conflicting opinions by discussing related evidence and arguments.''

\subsubsection{Minority opinion (subset of Consensus generator)} 
\label{sec:minority-opinion}

\textbf{In the minority opinion archetype, the LLM is called upon to give greater weight (or even to surface) underconsidered or overlooked perspectives.} It can be used in both group decision-making contexts (e.g, with explicit instructions to argue for the minority opinion \cite{lee_amplifying_2025}) or individual decision-making contexts (e.g., emphasize how the decision might be viewed from a minority perspective). This archetype is distinct from \textit{counterargument} because it is not necessarily eliciting the ``next best'' or next {statistically} likely perspective, and rather intentionally surfacing minority opinions (which could also potentially be considered fringe, incorrect, etc.)

{This archetype may require nuanced design to achieve the desired effect of promoting minority perspectives. For example, \citet{lee_amplifying_2025} aim to address power imbalances and increase ``psychological safety'' through their work, so their system design enables minority users to directly and privately share their opinions with a conversational AI, which then presents the minority user's opinion as the AI's own opinion to other majority users. At the same time, \citet{chiang_enhancing_2024} find that ``the LLM-powered devil’s advocate appears to have limited impacts on groups' appropriate utilization of AI assistance when its primary goal is to challenge the majority opinions within the group.'' They also \cite{chiang_enhancing_2024} reported a tendency for users to personify and anthropomorphize the ``LLM-powered devil's advocate'' during interactions, theorizing that this makes users feel challenged and less willing to engage in discussions. }

\subsubsection{Formalizer} 
The \textbf{formalizer human-LLM archetype aids in the construction of explicit formulations and/or optimization problems, including creating or translating problem constraints}. It is often asked to analyze a problem by text, and then generate variables, constraints, and formulas that can be used to solve the problem \cite{hao_planning_2025}. The LLM may be used to apply the formalized problem \cite{schoenegger_ai-augmented_2025}, or be integrated with other tools like Prolog for deductive reasoning \cite{borazjanizadeh_reliable_2024}. This approach could help bridge technical gaps between problem formalization and natural language descriptions of decision goals and constraints \cite{schoenegger_ai-augmented_2025}. At the same time, reported challenges with the \textit{formalizer} archetype included observed inaccuracies in LLM-suggested models \cite{wasserkrug_enhancing_2025} and high technical expertise requirements to debug errors within generated models \cite{lawless_i_2024}.

\subsubsection{Criteria applicator} The \textbf{human asks the LLM-as-criteria-applicator archetype to employ a specific set of instructions or guidelines to arrive at a decision}, which may or may not have been explicitly provided to the model. %
{This criteria was frequently deployed in medical domains, with an LLM asked to provide treatment recommendations according to guidelines developed by clinical professional societies. %
In some cases, the model was explicitly instructed to follow the relevant criteria, either by referencing the desired guidelines (e.g., \citet{sarangi_radiologic_2024} write prompt instructions to recommend procedures ``according to ACR Appropriateness Criteria'') or explicitly providing the guidelines in the prompt \cite{kaiser_interaction_2025}. Meanwhile, other papers evaluated LLM adherence to criteria while not explicitly providing or naming the guidelines \cite{buhr_assessment_2025}, including \citet{huo_performance_2024} indicating that their prompts did not contain ``reference to major surgical societies, organizations, or countries to mitigate bias.'' Model and design choices may be relevant towards the criteria archetype, as more specialized criteria may or may not be contained within the training corpus of various LLMs. }

{In addition to widespread observed use in diagnostic and treatment recommendation medical tasks, the \textit{criteria applicator} archetype may be used to facilitate automation of decision-making by processing data inputs according to instructed methods. For example, \citet{almekkawi_comparative_2025} explore multi-modal reasoning abilities of LLMs by estimating spino-pelvic parameters from a scoliosis radiograph. In a non-medical task, \citet{an_measuring_2025} employ LLMs to evaluate resumes for hiring decisions, also applying equal opportunity criteria towards the resume-based decision-making and finding evidence of race/gender bias.} %

\subsubsection{Judge} LLMs are increasingly being \textbf{used as a judge to rate or evaluate outputs (often generated at scale by LLMs themselves)}, a practice described as ``LLM-as-judge'' in the literature \cite{gu_survey_2025}. \textit{Judge} contrasts with the \textit{explainer} archetype, where the primary goal is to contextualize a decision by communicating with human, while \textit{judge} is primarily used to generate its own (generally quantitative) evaluation of output on some specific dimension. Metrics in our surveyed literature included political bias rankings \cite{agiza_politune_2025}, aggressiveness, accuracy \cite{gaber_evaluating_2025}, quality \cite{chen_recusersim_2025, azarbonyad_question-answer_2025, shaer_ai-augmented_2024}, and NLP-specific metrics like ROUGE and BLEU scores \cite{wang_can_2025}. The associated judge output could secondarily be used to communicate quality to a human. %
{Some papers found correlations between automatic \textit{judge} outputs and human judgments \cite{sandoval-castaneda_editduet_2025, shaer_ai-augmented_2024}, with \citet{wang_can_2025} finding that ``output-based methods with large state-of-the-art LLMs perform best'' in resembling human score patterns compared to other embedding- and probability-based \textit{judge} methodologies. }%

\subsubsection{Data processor} \textbf{In the data processor archetype, a human relies on LLM support for intermediate task steps rather than a final decision}, such as extracting unstructured data from structured text to support some other decision-making or registering human preferences for a given decision. This archetype is distinct from the \textit{judge} archetype as the primary goal is to shape or alter data (e.g., transforming a table to JSON format), rather than explicitly evaluate it. %
{The \textit{data processor} was often used to generate labels or annotations for raw text data \cite{ wang_human-llm_2024, yu_fine-tuning_2024, hota_evaluating_2024, nagaraj_rao_rideshare_2025, babatunde_moderating_2025}, which could be used for further analysis by the \textit{model} archetype or by another ML pipeline. \citet{zhang_llm-lade_2025} employ the archetype via `data augmentation,' using an LLM to transform semi-structured textual data logs into structured representations with standardized formatting. The archetype can also be used to shape data in other ways, including visually \cite{de_bari_evaluating_2024}.}
\begin{figure}[!h]
  \centering\includegraphics[width=1\linewidth]{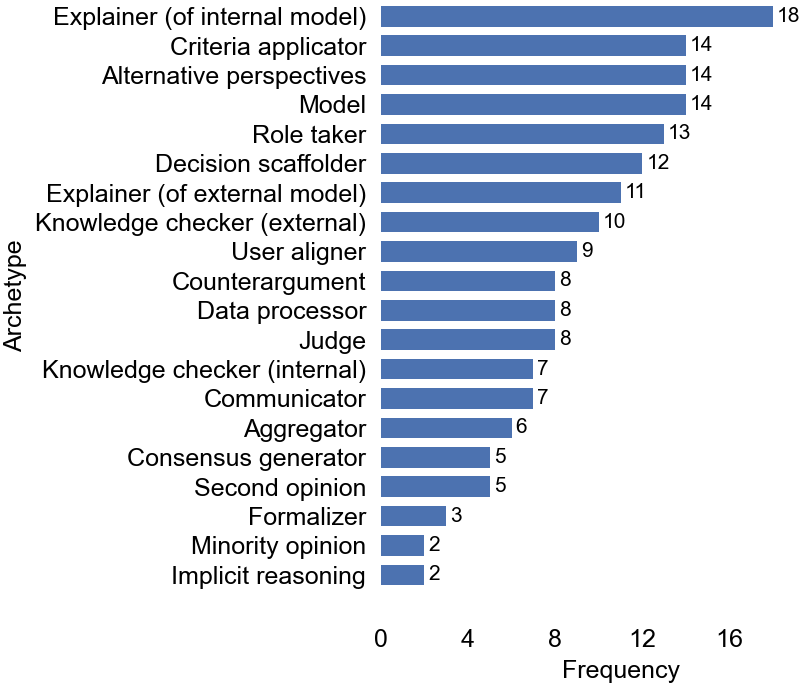}
   \caption{17 distinct archetypes were identified during scoping literature review of human-LLM decision-making, indicating the diverse socio-technical patterns used to engage humans and LLMs in shared reasoning and deliberation. 
   \textit{Explainer}, \textit{minority opinion}, and \textit{implicit reasoner} reflect subtypes of other archetypes.}
   \Description{Bar chart depicting the frequency distributions of the 17 archetypes identified in descending order. Explainer (of internal model) is the most frequent archetype with n=18 papers, while minority opinion and implicit reasoning are the lease frequent archetypes with n=2 papers. }
   \label{fig:archetype-frequencies}
\end{figure}

\subsection{Integrated and Combined Use of Archetypes}
\label{sec:switching}

{Though our 17 human-LLM archetypes are presented separately in \S\ref{sec:archetypes} for organizational and clarity, archetypes involve contextual and temporal dynamics that often enable integrated and/or simultaneous use. Archetypes may be used sequentially, in parallel, or selected ad hoc based on problem formulation, user preferences, or intended workflow design during decision-making. As such, we also analyzed archetype code groups across reviewed papers to identify patterns in how archetypes are used together. 

Of the 113 papers analyzed, we identified 48 papers (42\%) that employed or tested at least two archetypes, 11 papers with at least three archetypes, and three papers using at least four archetypes. Of the 48 papers employing multiple archetypes, the most frequently appearing archetypes were \textit{explainer (internal)} (n=14), followed by \textit{model} (13), \textit{alternative perspectives} (12), \textit{consensus generator} (8), and \textit{role taker} (7) (Figure \ref{fig:multiuse}a). This frequency pattern indicates that some archetypes may naturally align together (such as \textit{model} and \textit{internal explanations} of said models), while other archetype integrations may augment multi-step reasoning processes, including generation of initial or ``anchor'' predictions through \textit{model} or \textit{role taker} archetypes, followed by exploration into underconsidered or other plausible decisions via \textit{alternative perspectives} and \textit{counterargument} archetypes.

To further investigate common integrated archetype dynamics, we also analyzed the frequency of archetype pairs appearing across reviewed papers using multiple archetypes. Of the 48 papers combining archetypes, the most common pairs included \textit{model}--\textit{explainer (internal)} (n=8) and \textit{model}--\textit{data processor} (4), %
with the \textit{role taker} also frequently paired with other archetypes (Figure \ref{fig:multiuse}b). Some pairs indicate intuitive sequential patterns (e.g., use \textit{data processor} to shape raw data for further analysis by \textit{model} archetype, followed by a request for internal explanations \cite{zhang_llm-lade_2025}). Further, frequent use of \textit{role taker} with \textit{knowledge checker (internal)} and \textit{criteria applicator} reflect trends identified in \S\ref{sec:dimensions} where \textit{role taker} instructions are used to evaluate results within a specialized or expert domain.

Papers with more than three archetypes often entailed exploratory simulations of diverse human behavior, generating \textit{alternative perspectives}  \cite{lamparth_human_2025, sel_skin---game_2024} or \textit{counterarguments} \cite{ke_mitigating_2024}, and then implementing some \textit{consensus generation} to arrive at a final decision. Figure \ref{fig:multiuse}c illustrates varied approaches towards integrating archetypes across selected papers using at least 3 distinct archetypes. }

\subsubsection*{Summary}
Through our review of 113 research articles deploying LLMs within human-in-the-loop decision-making scenarios, we identified 17 different role archetypes that describe distinct human-LLM interaction patterns. 
While design factors like interface or prompt text are well-considered in the literature, our scoping review and mapping of human-LLM archetypes reveal important socio-technical patterns that govern how decisions are constructed and how roles are assigned to humans and LLMs, ultimately affecting decision-making processes and human-AI interactions.

System designers within the same decision-making task could employ LLM support through many diverse archetype strategies, with varying potential impacts on the decision-making process due to the design of the human-LLM interaction. 
To better understand the potential impacts or tradeoffs in adopting various human-LLM archetypes, we now consider a test case to translate various archetypes to a given real-world decision-making scenario, and examine relevant dimensions of LLM outputs that might affect downstream decision-making outcomes.

\begin{figure}
    \centering
    \begin{subfigure}[h]{\linewidth}
        \centering
        \includegraphics[width=\linewidth]{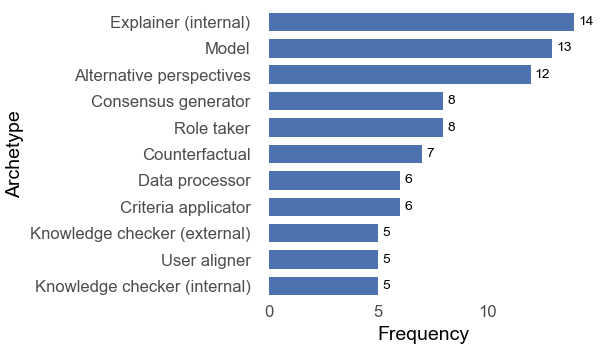}
        \caption{Archetypes most frequently integrated with other archetypes.}
         \Description{Most frequently integrated archetypes described in a bar chart containing 10 archetypes on the vertical axis. This bar chart displays ten archetypes on the vertical axis, beginning with "Explainer (Internal)” (most) and ending with “Knowledge checker (internal)” (lowest on the graph). The horizontal axis represents frequency, showing descending lengths of bars that indicate how often each archetype is observed in multiuse contexts.}
    \end{subfigure}

    \begin{subfigure}[h]{\linewidth}
        \centering
        \includegraphics[width=\linewidth]{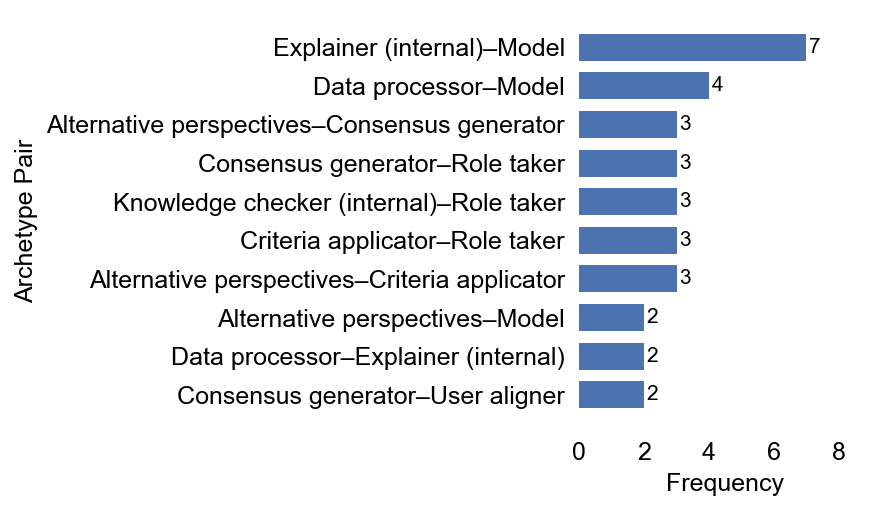}
        \caption{Archetype pairs appearing most frequently across reviewed papers.}
        \Description{Describes most frequently observed archetype pairs through a bar chart containing 10 linked archetypes pairs on the vertical axis. , pairs of archetypes, with each pair linked together. This bar chart displays ten archetypes on the vertical axis, beginning with “explainer (internal)-model” (most) and ending with “alternative perspectives-model” (lowest on the graph). The horizontal axis represents frequency, showing descending lengths of bars that indicate how often each archetype is observed in multiuse contexts.}
    \end{subfigure}
    
    \begin{subfigure}[h]{\linewidth}
        \centering
        \includegraphics[width=\linewidth]{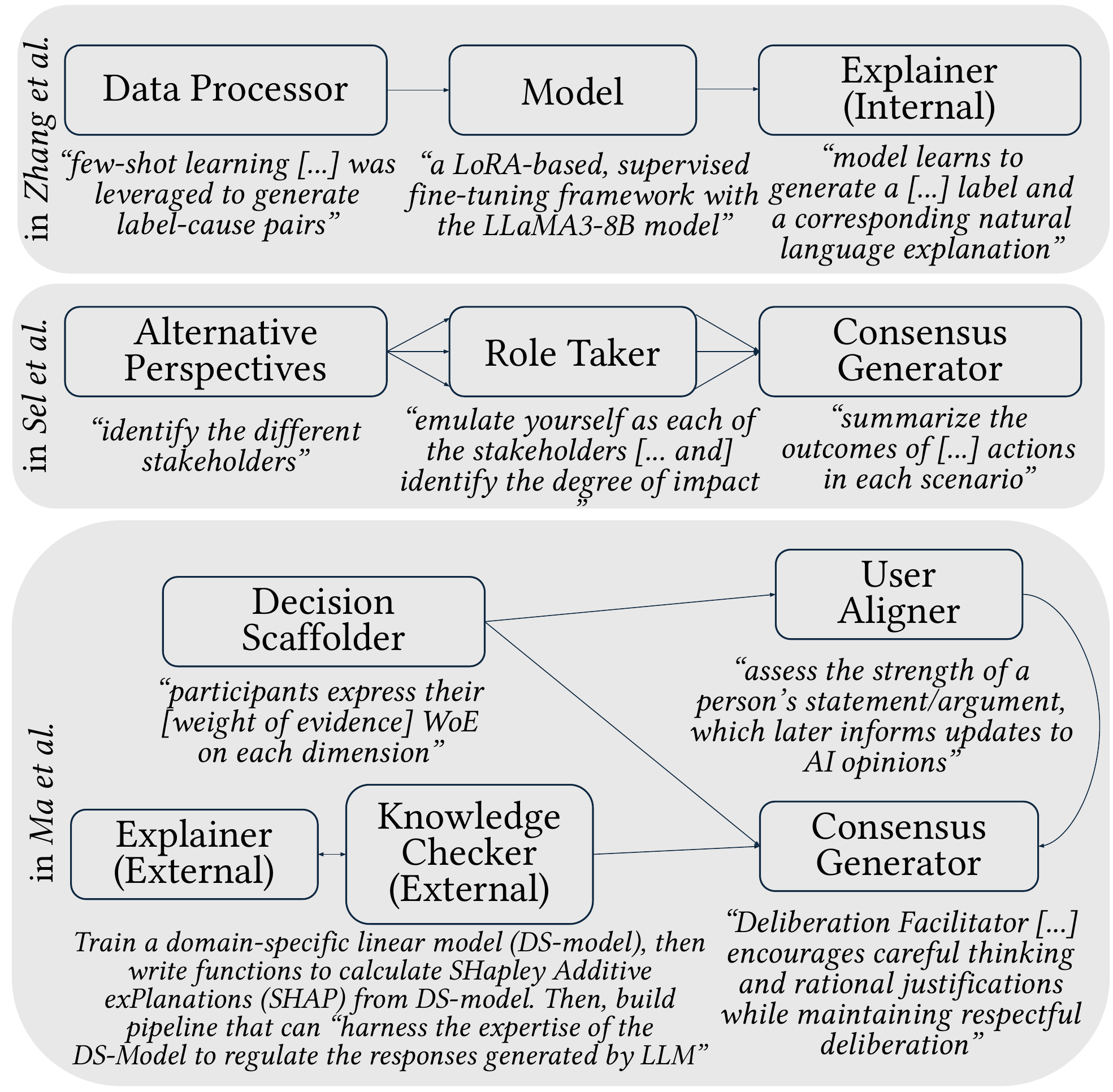}
        \caption{Example mappings from \citet{zhang_llm-lade_2025}, \citet{sel_skin---game_2024}, and \citet{ma_can_2025} of how multiple archetypes were used or integrated within a human-LLM decision-making workflow.}
        \Description{models integrated use of archetypes using flowchart diagrams. There are three separate floating sections with flowcharts representing three different papers. In the Zhang et al., example, data processor flows into model which flows into explainer (internal). In the Sel et al. example, alternative perspectives has three arrows that flow into role taker, and then three arrows flow out of role taker and flow into consensus generator. In the Ma et al. example, explainer (external) and knowledge checker (external) are linked together. Then, decision scaffolder flow into user aligner and consensus generator archetypes. The consensus generator also has links from the user aligned and the linked explainer/knowledge checker.}
    \end{subfigure}
   
    \caption{Archetypes can be combined, layered, and integrated within human-LLM decision-making contexts.}
    \Description{Three-part vertical figure exploring frequency and mappings of combined and integrated archetype use.}
    \label{fig:multiuse}
\end{figure}

\section{Clinical Case Study: Comparing Potential Effects of Archetypes Across Dimensions of Decision-making}
\label{sec:evaluation}

In \S\ref{sec:taxonomy}, we identified distinct human-LLM archetypes and related factors that distinguish human and LLM roles adopted during decision-making. While prior work emphasizes shaping of model behavior via instruction setting in system/user prompts, we highlight that problem framing and LLM role positioning are also very relevant during human-LLM decision-making and underexplored. 

Given that human-LLM archetypes are typically adopted and executed in isolated applications, there are few systematic comparisons exploring trade-offs in how alternative archetypes may perform within the same decision context. To bridge this gap, in this case study, we apply and compare multiple archetypes to a single high-stakes task and measure differences in key LLM output dimensions that could influence decision-making. Investigations into key factors including prediction accuracy, human-LLM agreement, and explanation quality are detailed separately in \S\ref{sec:results-accuracy} through \S\ref{sec:results-text-similarity}. 

By conducting preliminary analyses on direct LLM outputs, we 1) perform controlled comparisons across archetype choices, 2) quantify potential effects related to decision-making at scale, and 3) surface trade-offs that might emerge when builders of human-LLM systems select one archetype over another. %
These case study results can highlight potential tradeoffs between various archetypes on LLM-mediated decision-making, demonstrating the need for further attention, consideration, and inquiry into archetype selection and adoption.

\begin{figure*}
    \centering
    \includegraphics[width=1\linewidth]{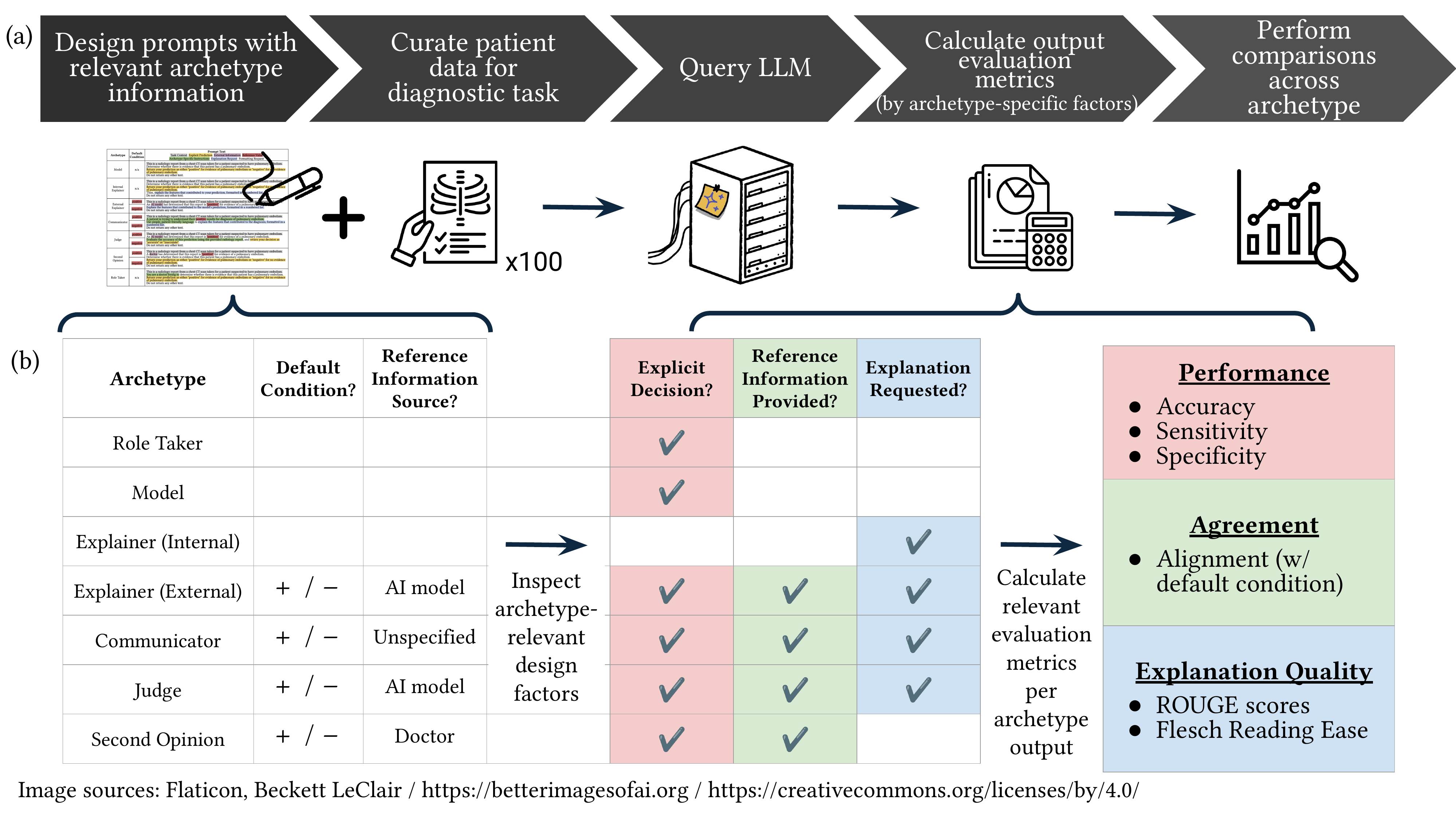}
    \caption{a) The case study followed an overall methodology where archetypes were translated to task-specific prompts, appended independently to 100 different patient reports, then queried to an LLM, with evaluation metrics calculated and compared across each archetype. b) The translation of archetypes to prompts often required the addition of reference information sources (details about who and what external predictions were made) and variations in default conditions (whether the reference prediction was positive or negative). These details shaped which evaluation metrics were calculated and compared across each archetype.    %
    }
    \Description{Two-part figure detailing the case study experimental design. The top portion (Figure A) displays a 5-step flowchart from left to right with descriptions of each step and icons depicting the step below. The steps are design prompts, curate patient data, query LLM, calculate output evaluation outcomes (based on archetype-relevant factors), and perform comparisons across archetypes.
    The bottom portion (Figure B) expands details of steps in the Figure A flowchart. The “design prompts” step is linked to a table describing the archetypes tested (including information on reference sources and whether default conditions are present). The “calculate output evaluation outcomes” step is linked to two color-coded tables. The first table details which archetypes have explicit decisions, reference information provided, and explanations requested. Then, an arrow links this first table to a second table detailing the different evaluation metrics used in the case study. The first table’s explicit decision column and second table’s performance cells are both colored red to indicate their relation. The first table’s reference information provided column and second table’s agreement cells are both colored green to indicate their relation. The first table’s explanation requested column and explanation quality cells are both colored blue to indicate their relationship.

}
    \label{fig:case-study-methods}
\end{figure*}

\subsection{Task} 
\label{sec:task}

To facilitate a comprehensive comparison of archetypes, we focused on a complex decision-making scenario likely to accommodate diverse problem-solving strategies. While many question-answering benchmarks exist to evaluate LLM outputs, we sought a task rooted in real-world data. 
Thus, we selected a clinical decision-making task involving diagnosis from real-world radiology reports. The medical domain was chosen due to its high-stakes nature, where decisions critically impact patient health and often require subjective, complex, and collaborative input from experts. %
Many archetypes reviewed in \S\ref{sec:archetypes} were applied to diagnostic and treatment recommendation tasks, demonstrating that these archetypes are used in isolation, but often not systematically compared or evaluated in comparison. 

By applying select archetypes identified in \S\ref{sec:taxonomy} to this case study, we simulate diverse human-LLM decision-making contexts and enable controlled comparison of LLM outputs across key dimensions influencing decision-making, including accuracy, agreement, and explanation quality. 
{Since these LLM outputs would be used by a medical expert in a human-in-the-loop (HITL) decision making framework, these pilot quantitative results can indicate relevant foundations for future qualitative investigations into the influence of human-LLM archetype selection. }

Specifically, this section is structured as follows:
In \S\ref{sec:task}, we discuss details of our real-world task. More specifically, in \S\ref{sec:report-data}, we describe the real-world dataset employed to obtain test cases. In \S\ref{sec:prompt-translation}, we describe how the archetypes mapped in \S\ref{sec:archetypes} were translated and applied to the diagnostic decision-making task for controlled comparison. Then, \S\ref{sec:implementation} describes how the real-world test cases and archetype translations were combined as queries to an LLM. 
In \S\ref{sec:results-accuracy}-\ref{sec:results-text-similarity}, we detail the tradeoffs observed across key decision-making dimensions influenced by various archetypes, with relevant methodology, metrics, and takeaways discussed in each section. In all, these test explorations across various LLM outputs demonstrate the emerging potential for various archetypes to impact decision-making.

\subsubsection{Data}
\label{sec:report-data}

We based our task on radiology reports and ground truth labels curated from the publicly available INSPECT EHR dataset of pulmonary embolism (PE) patients treated at Stanford Hospital \cite{huang_inspect_2023}. Specifically, the task involves evaluating a radiology report of a computer tomography pulmonary angiogram (CTPA) for evidence of pulmonary embolism.

After applying for credentialed access to the INSPECT EHR database, we obtained CPTA radiology reports (n=23,248) for patients suspected to have a PE.
Then, we pre-processed reports by linking them to patient IDs and manually derived ground truth labels (20\% positive cases) from \cite{banerjee_comparative_2019}, and standardizing text case. %
We excluded reports with case-insensitive matches to ``embol*'' to remove straightforward cases where the evidence was clearly confirmed (e.g.,  ``no evidence of pulmonary embolism'' or ``saddle emboli present''). This allowed us to focus on borderline and edge cases, reflecting real-world instances requiring consultation from additional colleagues or tools to aid decision-making in complex scenarios. 

After filtering for these complex reports, our dataset included 2915 negative and 167 positive ground truth PE cases across 2894 unique patients (5\% positive cases). %
We randomly sampled a subset of 50 negative and 50 positive reports for our test case (total n=100). Positive reports were oversampled to test the discriminative ability of archetypes in this high-stakes medical scenario, where a false negative could leave a patient with a serious condition untreated. This approach also allowed us to assess the impact of modifying default reference or anchor information across relevant archetype prompts. 

\subsubsection{Translating archetypes to a task.}
\label{sec:prompt-translation}

Given the variety of methods and prompts availble for executing various archetype, we created a prompting framework that systematically perturbed {key} variables distinguishing archetypes identified in \S\ref{sec:archetypes}. These distinguishing variables included LLM role, {reference information and decision social context (context on AI model vs. expert opinions)}, and problem solving instructions. We conducted systematic comparison across outputs from different archetype conditions after varying and/or standardizing information contained within the following ``prompt building blocks:'' 

\begin{itemize}
    \item task context
    \item external information (including social role)
    \item default condition (e.g., reference information that an external judgment was `positive' or `negative' for this case) 
    \item archetype-specific details
    \item explicit prediction 
    \item explanation request
    \item formatting instructions
\end{itemize}

The exact prompts and associated perturbations for each archetype can be identified in Figure \ref{fig:experiment-prompts}. This prompt perturbation paradigm was constructed to enable systematic comparison of the potential effects of adopting various archetypes across a single decision-making task.

\subsubsection{Implementation. }
\label{sec:implementation}

{The overall case study evaluation workflow is described in Figure \ref{fig:case-study-methods}a. Each archetype prompt authored in \S\ref{sec:prompt-translation} was independently appended to each report in the common set of 100 radiology reports sampled in \S\ref{sec:report-data}, then queried to the LLM. As such, 100 LLM queries were performed for each archetype-report pair, with 100 outputs analyzed across each unique archetype (and unique default condition where relevant).}

We conducted our archetype prompt test cases using the OpenAI GPT model series given its widespread usage identified in \S\ref{sec:archetypes}. The analyses required privacy-compliant frameworks that do not store queries or outputs to comply with data usage agreements \cite{huang_inspect_2023} and appropriately replicate likely real-world usage patterns. As a result, queries were sent to GPT-4o (FedRAMP-accredited model, temperature parameter set to 0) using Azure Government OpenAI endpoints, with data not retained or used for retraining and accessed securely for HIPAA compliance. 

Each independent query appended an archetype prompt to a de-identified radiology report. Then, resulting outputs were stored in a table, with further output text analyses and comparisons performed using Python 3.12 in a Jupyter notebook. {Then, we calculated relevant evaluation metrics across LLM outputs (described further in \S\ref{sec:results-accuracy}-\ref{sec:results-text-similarity}), using archetype-specific attributes to select relevant metrics for comparison (Figure \ref{fig:case-study-methods}b).}

\subsection{Archetypes that elicit explicit predictions from LLMs can affect model decision accuracy.} 
\label{sec:results-accuracy}

First, we examined the potential effects of human-LLM interaction archetypes on accuracy of LLM-generated outputs. Since accuracy calculations involve comparison of a prediction to ground truth, these tests examined all archetypes that explicitly elicit a prediction from the LLM [\textit{Role Taker, Model, Judge, Internal Explainer, Second Opinion}], with exact prompts detailed in Figure \ref{fig:experiment-prompts}. This performance analysis aimed to understand how decision-making instructions and context provided by various archetypes influence the distribution of predictions made by an LLM. 

\subsubsection{Methods} 

Archetypes that elicit explicit binary judgments (e.g., positive/negative, accurate/inaccurate) were translated into prompts with common elements (task context, request for explicit prediction, and formatting instruction), and systematically perturbed across archetype-specific attributes(external information, default conditions, and explanation requests (where appropriate) as detailed in \S\ref{sec:prompt-translation}. Each archetype prompt was appended to all 100 radiology reports sampled in \S\ref{sec:report-data} and queried independently to GPT-4o per \S\ref{sec:implementation}. Then, binary predictions were extracted using string matching methods applied to LLM output text, compared to the ground truth labels obtained manually by \cite{banerjee_comparative_2019}, and used to calculate accuracy, specificity, and sensitivity.

To assess statistically significant differences in archetype-related LLM prediction distributions, we performed Cochran's Q test across all explicit model predictions and used McNemar tests for pairwise comparisons between archetype predictions. {These tests were employed to evidence the potential variable effects of adopting different archetypes (rather than to evaluate performance on the task itself).}

\subsubsection{Results.}

\begin{table}[]
\begin{tabular}{>{\centering\arraybackslash}p{0.4\linewidth}
>{\centering\arraybackslash}p{0.13\linewidth}
>{\centering\arraybackslash}p{0.15\linewidth}
>{\centering\arraybackslash}p{0.15\linewidth}}
    \toprule
    \textbf{Archetype}        &  \textbf{Accuracy} &  \textbf{Sensitivity} &  \textbf{Specificity} \\ \midrule
    Explainer (internal)      & 0.93              & 0.86                 & 1.00                 \\
    Model                     & 0.93              & 0.88                 & 0.98                 \\
    Judge (negative)          & 0.87              & 0.88                 & 0.86                 \\
    Judge (positive)          & 0.95              & 0.94                 & 0.96                 \\
    Second Opinion (negative) & 0.90              & 0.80                 & 1.00                 \\
    Second Opinion (positive) & 0.94              & 0.90                 & 0.98                 \\
    Role Taker                & 0.94              & 0.88                 & 1.00      \\  
    \bottomrule
\end{tabular}
\caption{Differences in performance metrics, including accuracy, sensitivity, and specificity, can be observed across archetypes that elicit explicit predictions from an LLM.}
\label{table:accuracy}
\Description{A seven-column by seven-row table reporting pairwise McNemar test p-values for comparisons between seven explicit-prediction archetypes. The column and row headers list the archetypes in the same order: Explainer (Internal), Model, Judge (Negative), Judge (Positive), Second Opinion (Negative), Second Opinion (Positive), and Role Taker. Each cell at the intersection of a row and column gives the p-value for that pairwise comparison; diagonal cells (an archetype compared with itself) are left blank or marked with a dash. Several cells include an asterisk indicating p < .05.}
\end{table}

\begin{table*}[h!]
\begin{tabular}{@{}cccccccc@{}}
\toprule
\textbf{} &
  \begin{tabular}[c]{@{}c@{}}Explainer\\ (Internal)\end{tabular} &
  Model &
  \begin{tabular}[c]{@{}c@{}}Judge\\ (Negative)\end{tabular} &
  \begin{tabular}[c]{@{}c@{}}Judge\\ (Positive)\end{tabular} &
  \begin{tabular}[c]{@{}c@{}}Second Opinion\\ (Negative)\end{tabular} &
  \begin{tabular}[c]{@{}c@{}}Second Opinion\\ (Positive)\end{tabular} &
    \begin{tabular}[c]{@{}c@{}}Role\\ Taker\end{tabular}  \\ \midrule
Explainer (Internal)       & -         &          &            &           &          &     &   \\
Model                     & 0.5       & -        &            &           &          &     &   \\
Judge (Negative)           & 0.0078* & 0.031* & -          &           &          &     &   \\
Judge (Positive)           & 0.070   & 0.22   & 0.77     & -         &          &     &   \\
Second Opinion (negative) & 0.25      & 0.0625   & 0.00098* & 0.0039* & -        &     &   \\
Second Opinion (positive) & 0.25      & 1        & 0.125      & 0.375     & 0.031* & -   &   \\
Role Taker               & 1         & 1        & 0.016*   & 0.125     & 0.125    & 0.5 & - \\ \bottomrule
\end{tabular}
    \caption{p-values from McNemar tests for pairwise comparisons across explicit predictions archetypes, with asterisks denoting $p<.05$.} 
    \label{table:performance-comparisons}
\end{table*}

All models demonstrated relatively high levels of accuracy in their predictions, with \textit{Judge (positive)} and \textit{Judge (negative)} reporting the highest and lowest accuracy rates, respectively (Table \ref{table:accuracy}). The sensitivity and specificity rates for the \textit{Judge (positive)} archetype were particularly high. Many archetypes, including the \textit{Model, Internal Explainer, Role Taker,} and \textit{Second Opinion} exhibited or near perfect specificity, indicating that reductions in accuracy were from false negatives (e.g., failing to correctly categorize a report with evidence of PE).  

Cochran's Q test indicated statistically significant differences between the binary prediction distributions of the 6 tested archetypes [\textit{Internal Explainer}, \textit{Model}, \textit{Judge} (negative and positive), Second Opinion\textit{} (negative and positive), and \textit{Role Taker}] ($p$=\num{3.75e-05}). Pairwise comparisons are reported in \S\ref{sec:case-study-agreement} and Table \ref{table:performance-comparisons}. Thus, our \textbf{results indicated statistically significant differences in prediction accuracy between archetypes that explicitly elicited a prediction from the model}.

\OvalColumnBox{
\textit{Summary.} Accuracy comparisons show that archetype choices can shift LLM prediction distributions and measurably change performance tied to archetype-specific cues (e.g., reference roles and values). }

\subsection{Archetypes receiving external reference information may  display varying levels of agreement.}
\label{sec:case-study-agreement}

Given concerns over sycophancy or agreeableness of LLMs \cite{pan_user-assistant_2025}, we also sought to evaluate whether various archetypes might affect agreement between LLMs and the decision information referenced in prompts.

\subsubsection{Methods.} Employing a similar framework to \S\ref{sec:results-accuracy}, we identified archetypes that archetypes that provide explicit judgments after receiving external reference information [\textit{Judge, External Explainer, Second Opinion}].\footnote{The \textit{Counterargument} archetype was not included because it explicitly asks for a different prediction, eliminating agreement.} Then, we crafted prompts that systematically perturbed information related to the archetype including the origin of the external information (doctor for \textit{Second Opinion} or AI model for \textit{External Explainer}), and the default/anchor condition (positive or negative prediction). 

Agreement levels were calculated between archetypes providing reference or external judgments. Agreement was calculated as the proportion of LLM output responses aligning with the reference judgment provided (e.g., the LLM predicting `positive' when the \textit{Second Opinion} archetype prompt provided a reference doctor decision of `negative' was noted as a disagreement).

\subsubsection{Results.}

There were 6 report cases in which the \textit{second opinion archetype} agreed with the prompts providing a reference doctor judgment of both `positive' and `negative.' McNemar tests were conducted to further compare differences in prediction distributions between archetypes tested in \S\ref{sec:results-accuracy}. The binomial p-value matrices for these comparisons are reported in Table \ref{table:performance-comparisons}.

Results from pairwise comparisons across explicit prediction archetypes indicated statistically significant differences between the second opinion archetypes ($p=0.03125$), but not the judge archetypes ($p=.7744$). The higher agreement rate in the \textit{second opinion} (negative) outputs indicated that the tested LLM was more likely to defer to a doctor when it reported evidence as absent, potentially creating more false negatives. 

The negative-reference second opinion archetype also produced statistically significant differences in LLM predictions compared to the judge archetype (with both reference predictions). As such, swapping the reference source (e.g., AI model versus doctor) produced observable shifts in LLM judgment distributions. The reference value of the second opinion archetype may also be relevant, as swapping the doctor's reported judgment between positive and negative judgments caused statistically significant shifts.

\OvalColumnBox{
\textit{Summary.} Agreement with external judgments can be influenced by archetype-specific cues including the reference judgment (e.g., positive vs. negative prediction) and who provides it (e.g., AI model or human expert). 

}

\begin{table}[!hbt]
\begin{tabular}{>{\centering\arraybackslash}p{0.45\linewidth}
>{\centering\arraybackslash}p{0.2\linewidth}
>{\centering\arraybackslash}p{0.2\linewidth}}
    \toprule
    \textbf{Archetype}        & \textbf{Agreement Value} & \textbf{Agreement Rate} \\ \midrule
    Judge (negative)          & ``accurate''                 & 0.49                    \\
    Judge (positive)          & ``accurate''                 & 0.49                    \\
    Second Opinion (negative) & ``negative''                 & 0.46                    \\
    Second Opinion (positive) & ``positive''                 & 0.60    \\    
    \bottomrule
\end{tabular}
\caption{The {Agreement Value} column indicates the string output returned by the LLM that would indicate agreement with reference information provided in the prompt (e.g., evaluating an AI model's prediction as `accurate' indicates agreement). Agreement rates represent the proportion of LLM outputs that matched the agreement value.} %
\label{table:agreement}
\Description{A three-column table with four data rows. The columns include Archetype (short label of archetype name), Agreement Value (the string output returned by the LLM that would indicate agreement with reference information provided in the prompt), and Agreement rate (proportion of LLM outputs that matched the agreement value out of 1.0). The archetypes included are judge (negative), judge (positive), second opinion (negative), second opinion (positive).}
\end{table}

\subsection{Archetypes may facilitate LLM explanations of varying complexity but similar text similarity.}\label{sec:results-text-similarity}

Next, we sought to understand the effect of archetypes on explanations that may be generated for human users. 

\subsubsection{Methods.}

We compared LLM outputs across archetypes that elicit explanations for users [\textit{Communicator}, \textit{Internal Explainer}, and \textit{External Explainer}] described in Figure \ref{fig:experiment-prompts}.] ROUGE text similarity scores \cite{lin_rouge_2004} were calculated (using the rouge\_score Python package) across outputs with explanations and using the original report as the reference text. These metrics are similarly employed in other studies for quantitative evaluation of explanations derived from text \cite{kim_medexqa_2024, guo_automated_2022, li_joint_nodate}. 
Wilcoxon-signed-rank tests (with Bonferroni correction) were applied to conduct pairwise comparisons across ROUGE scores calculated for archetypes explaining opposing predictions (e.g. comparisons of positive vs. negative explanations within archetypes).

Flesch Reading Ease readability scores were also calculated (using the textstat Python package) for all text outputs. This readability score has been employed as a proxy metric for text accessibility and cognitive load to evaluate LLM as communicators in similar studies \cite{davis_evaluating_2023, kattih_artificial_2024}).

\subsubsection{Results.}

\begin{table}[!b]
\begin{tabular}{@{}>{\centering\arraybackslash}p{0.45\linewidth}
>{\centering\arraybackslash}p{0.15\linewidth}
>{\centering\arraybackslash}p{0.15\linewidth}
>{\centering\arraybackslash}p{0.15\linewidth}@{}}
    \toprule

    \textbf{Archetype} & \small \textbf{ROUGE-1} & \small \textbf{ROUGE-2} & \small \textbf{ROUGE-L} \\
    \midrule
    \normalsize
    Communicator (positive)       & 0.34 & 0.11 & 0.24 \\
    Communicator (negative)       & 0.35 & 0.11 & 0.24 \\
    External Explainer (positive) & 0.49 & 0.34 & 0.41 \\
    External Explainer (negative) & 0.49 & 0.28 & 0.37 \\
    Internal Explainer            & 0.42 & 0.22 & 0.31 \\
    \bottomrule
    
\end{tabular}

    \caption{ROUGE text similarity scores calculated across archetypes eliciting explanations.}
    \label{fig:text-similarity}
    \Description{A four-column table with five data rows. The columns are Archetype (short label of archetype name), ROUGE-1 (unigram overlap score out of 1.0), ROUGE-2 (bigram overlap score out of 1.0), and ROUGE-L (longest common subsequence score out of 1.0). The rows report ROUGE scores for five archetypes: Communicator (positive), Communicator (negative), External Explainer (positive), External Explainer (negative), and Internal Explainer.}
\end{table}

Wilcoxon-signed-rank tests indicated that there were no statistically significant differences in text similarity between the positive and negative communicator archetypes across all three metrics (Rouge-1 p=1, Rouge-2 p=1, Rouge-L p=1). There were also no statistically significant differences between the Rouge-1 (p=1) and Rouge-L (p=.698) scores for the negative and positive external explainer archetypes, but significant differences identified for the Rouge-2 (p=.033). 

These results could hint towards two possibilities. First, models may use similar evidence from the same report to support opposing judgments. This suggests that LLMs may not be well equipped to discriminate or push back on accuracy when instructed to give explanations supporting (incorrect) decisions, indicating behaviors of deference. Second, text similarity scores may not be adequate for evaluating explanation quality, as they may not account for an overall message or argument. %

To explore these possibilities, we qualitatively compared outputs between archetypes instructed to give explanations supporting a positive and negative prediction for a given report. We observed instances where the model explanation instead highlighted potential errors contributing to the prediction it was instructed to support (e.g., an \textit{External Explainer} asked to incorrectly support a positive prediction noted ``the phrase `no evidence of residual or recurrent disease' may have been misinterpreted by the model as indicating the absence of prior embolism rather than the current absence of embolism''). 
We also noted cases where the model's explanations contradicted the overall judgement (e.g., ``1. the scan did not show any signs of new or remaining blood clots in the lungs (no evidence of pulmonary embolism)'' for a positive result \textit{Communicator}). Lastly, high similarity in text between explanations for opposing decisions often indicated subjective interpretations of the same evidence. For example, an \textit{External Explainer} prompted to explain different predictions for the same report returned ``presence of a small intraluminal filling defect within the pulmonary artery supplying the right lower lobe, indicative of a pulmonary embolism'' for the positive prediction, and ``presence of a small intraluminal filling defect, which may not meet the threshold for a definitive pulmonary embolism diagnosis'' for \textit{External Explainer} for the negative prediction.

\begin{table}[!t]
\begin{tabular}{@{}c
>{\centering\arraybackslash}p{0.15\linewidth}
>{\centering\arraybackslash}p{0.3\linewidth}@{}}
    \toprule
    {\textbf{Archetype}} & \textbf{Readability}        & \textbf{Text Length (words)} \\ \midrule
    Communicator (positive)        & 61.4  & 133      \\
    Communicator (negative)        & 59.0  & 143      \\
    External Explainer (positive)  & 19.8 & 50      \\
    External Explainer (negative)  & 12.9 & 80      \\
    Internal Explainer       & 13.1 & 64      \\ \bottomrule
    \end{tabular}
    \caption{Average readability scores and text length for explanations provided across LLM archetypes. Note that the readability score captures context including sentence length (in words), but not overall text length.} 
    \label{fig:readability}
    \Description{A three-column table with five data rows. The columns are Archetype (short label of archetype name), Readability (average readability score), and Text Length (average number of words). The rows report values for five archetypes: Communicator (positive), Communicator (negative), External Explainer (positive), External Explainer (negative), and Internal Explainer.}
\end{table}

The \textit{Communicator} archetype scored much higher than the \textit{External Explainer} on readability scores. This result is expected given that the \textit{Communicator} archetype prompt explicitly included instructions to use patient-friendly language. While the \textit{Communicator} outputs displayed similar average readability scores and lengths, the external explanations for negative predictions displayed lower readability scores and longer explanations, potentially relating to higher cognitive load \cite{davis_evaluating_2023, kattih_artificial_2024}.

\OvalColumnBox{
\textit{Summary.} Archetypes that elicit explanations to support user deliberation can influence output complexity and readability (potentially affecting users' cognitive load). %
LLMs can produce explanations for both accurate and inaccurate predictions, and explanations for opposing predictions can be textually similar but vary in complexity. Further work is needed to examine the effectiveness, accuracy, and usefulness of LLM-generated explanations. 

}

\subsection{Limitations}

We recognize that archetypes are defined to reflect complex socio-technical environments that ultimately shape the design of human-LLM systems. To enable controlled comparison between archetypes, the cases analyzed in \S\ref{sec:evaluation} only tested variations in user text and patient data inputs {and presented quantitative evaluations of the outputs.} The prompts and task selected only reflect one plausible strategy to implement and test differences in human-LLM archetypal decision-making strategies. Our results indicated that archetypes have actual effects on the output dimensions we quantitatively analyzed. As such, we demonstrate that archetypes can have impacts, and thus there is a need for real-world consideration of archetypes -- in this way, our analysis can lay foundations for future investigations into how users and decision-makers prefer or rely on various archetypes and how they can be better and more appropriately designed. 

While many works surveyed in \S\ref{sec:archetypes} engaged zero-shot frameworks without fine-tuning (including all those in \textit{knowledge checker (internal)} and many under \textit{criteria applicator}), developers of human-LLM systems may seek to perform fine-tuning operations prior to testing and deployment. As such, designers should consider how human-LLM archetypes might impact model performance and decision-making dynamics within their particular model, task, and system design contexts.

Our archetype test cases were only performed on one model, with variations in prompts test across a subset of identified archetypes. We acknowledge that the tests conducted in \S\ref{sec:evaluation} are not meant to characterize the behavior of GPT-4 or advocate for a specific archetype based on task performance, whose outputs and parameters will inevitably change across model updates and re-training. Rather, the presented tests and results seek to probe the archetypes in a real-world context against each other, and demonstrate potential tradeoffs or impacts from adopting certain archetypes. 

Lastly, we also acknowledge that output explanations in \S\ref{sec:results-text-similarity} were evaluated using automated text similarity metrics, which may not finely capture inaccuracies or subjective quality to humans. However, these metrics are routinely used in NLP evaluations and enabled comparison across hundreds of LLM outputs, which could not be feasibly evaluated by human experts. Future work may engage expert clinicians during in-depth interviews to understand their subjective use and reliance of various human-LLM archetypes and their associated outputs.

\section{Design Choice Dimensions of Archetypes}\label{sec:dimensions}

Our findings from \S\ref{sec:archetypes} revealed various archetypes that LLMs occupy during HITL decision-making paradigms. Through thematic analysis described in \S\ref{sec:taxonomy}, we inductively identified several dimensions affecting the selection, use, and downstream effects of human-LLM archetypes. Our case study in \S\ref{sec:evaluation} also demonstrated that archetypes affect real-world dimensions affecting task outcomes, including accuracy, agreement, and explanation quality. 

{In this section, we synthesize patterns in human-LLM archetype interactions from our thematic analysis described in \S\ref{sec:taxonomy}, exploratory comparisons of archetypes effects in \S\ref{sec:evaluation}, and broader critical computing literature. We identify seven critical dimensions of human-LLM archetypes that require consideration from designers of HAI decision-making systems. These seven dimensions reflect decision points involving tradeoffs and requiring cautious analysis prior to adoption and during evaluation, including where combined use of archetypes increases human-LLM system complexity.} 

\subsection{Decision control: From AI to human autonomy} 

\begin{figure*}[!hbt]
    \centering
    \includegraphics[width=1\linewidth]{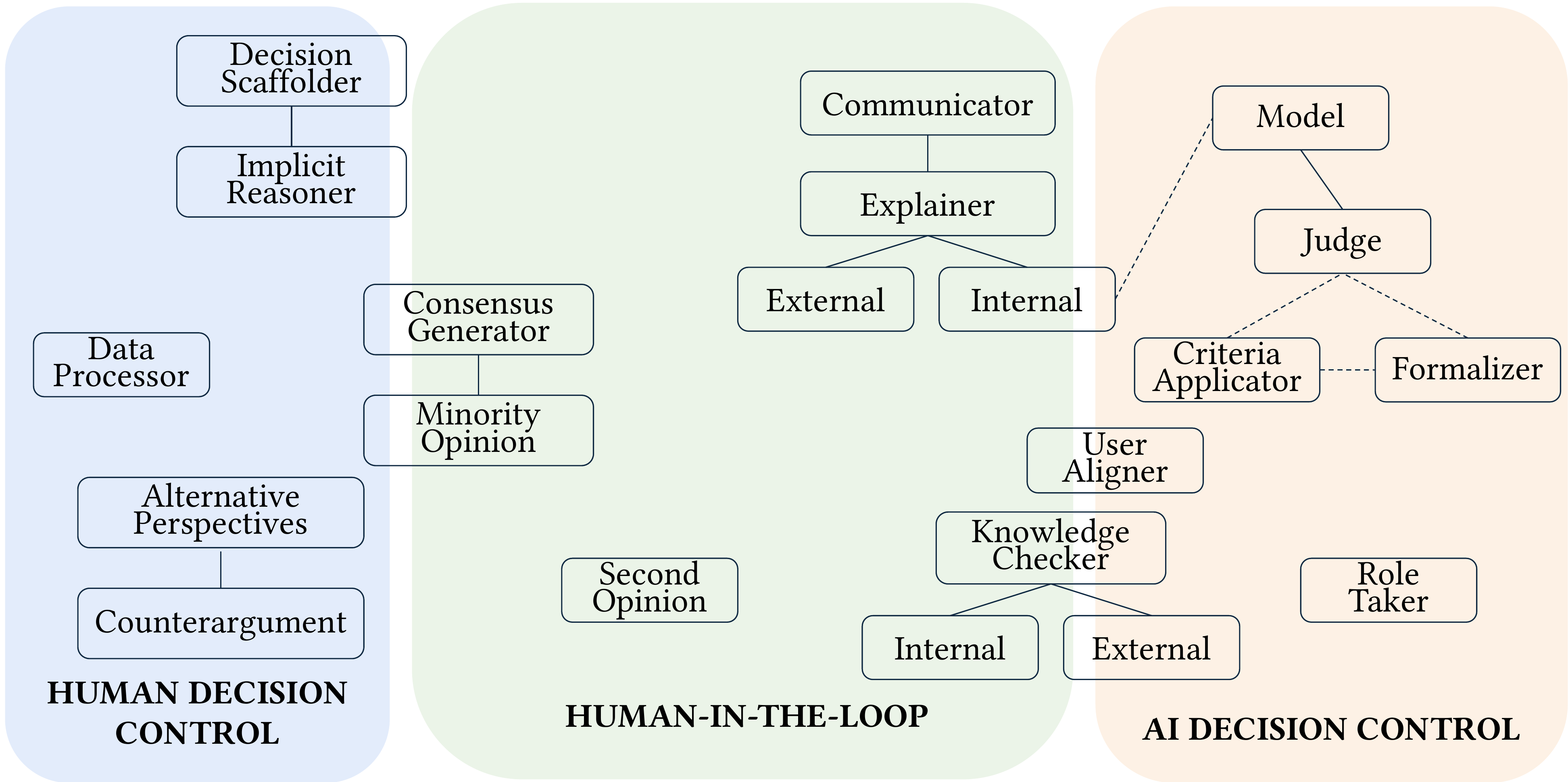}
    \caption{The selection of human-LLM archetypes involves assignment of varying autonomy levels to human and AI decision-makers. Solid lines demonstrate archetypes that involve direct subtypes, while dashed lines indicate archetypes that may be related or deployed simultaneously.}
    \label{fig:decision-control}
    \Description{Figure mapping all 17 archetypes in one of three categories. In the first category of human decision control, the data processor, decision scaffolder, implicit reasoner, alternative perspectives, and counterfactual archetypes are present. In the human-in-the-loop category, the consensus generator, minority opinion, second opinion, communicator, explainer, and knowledge checker archetypes are present. In the AI decision control category, model, judge, criteria applicator, formalizer, user aligner, and role taker archetypes are present.}
\end{figure*}

Human-AI systems can be designed with varying levels of control given to the humans and/or AI systems involved in decision-making \cite{faggioli_perspectives_2023}. {Each archetype cedes or invokes various levels of LLM control within the decision framework, illustrated in Figure \ref{fig:decision-control} where all 17 archetypes are positioned on a spectrum of decision control ranging from human autonomy. Most \textit{role taker / [persona} and \textit{judge} archetypes elicit a final decision or prediction from the LLM, enabling human deference or automation bias \cite{schaffer_i_2019}. Meanwhile, many HITL archetypes include varying ranges of human/AI control, such as the \textit{consensus generator} synthesizing human judgments already offered, or the \textit{Communicator} and \textit{Knowledge Checker} archetypes giving natural language outputs to be interpreted and potentially acted on by a human.} Some archetypes like \textit{data processor} and \textit{implicit reasoner} reflect higher human decision control because they conduct intermediate steps for human consideration during deliberation.{ Yet, as observed in \S\ref{sec:switching}, these higher human autonomy archetypes may be combined with other more automated archetypes, including \textit{model} or \textit{judge}. Data cleaning, processing, and structuring are also value-laden steps that can influence downstream algorithmic outcomes \cite{chappidi_manual_2025, agniel_biases_2018, sambasivan_everyone_2021, geiger_garbage_2020, simson_lazy_2024, gondimalla_aligning_2024, pushkarna_data_2022}. }%
As such, those designing systems for decision-making should consider their desired levels of human-AI decision control and accompanying archetypes, {including how autonomy may be shifted across integration of various archetypes.}

High LLM decision autonomy may raise concerns across multiple domains. Emerging research shows that the framing of a problem within a prompt is heavily relevant when directly prompting LLMs for decisions (e.g., asking LLMs for an action vs. an omission produces different results in moral dilemmas \cite{cheung_large_2025}). Our \S\ref{sec:results-accuracy} case study explorations of archetypes with higher AI autonomy observed that model performance was influenced by reference judgments and other prompt details. These results suggest that error analysis may relevant during archetype selection, as models displayed varying performance across sensitivity and specificity metrics. For example, while the \textit{role taker} archetype is popular and a predominant approach to LLM-supported decision-making, it displayed a relatively lower sensitivity compared to other archetypes which may not be preferable in a diagnostic context with high costs for false negatives. %
However, given the non-deterministic nature of model outputs and high variability across models, further research should explore prediction performance across archetypes in diverse contexts.

Decision control also brings legal implications of who remains responsible for outcomes when AI is consulted or given autonomy during decision making. While this paper focused specifically on HITL applications of LLMs, some archetypes could inadvertently (or intentionally) enable autonomous decision-making facilitated due to automation bias or other factors. Art. 22 of the EU GDPR, for example, concerns fully automated decision-making \cite{noauthor_art_nodate}, bringing governance and regulatory considerations for more autonomous archetypes. 

    
\OvalColumnBox{\textit{Summary.} {{Designers of human-LLM systems should consider their target levels of AI autonomy and examine whether their chosen archetype(s) enables this strategy.}} For example, while popular and easy to implement, the \textit{Role Taker\slash Persona} archetype, is related to high levels of AI autonomy and may not be desired (or permissible) in many domains.}

\subsection{Internal vs. External Knowledge} 

Some archetypes facilitate decision-making by presenting natural language reasoning or knowledge obtained from large-scale training of a model (e.g., \textit{internal explainer}), while others attempt to provide specialized knowledge obtained via retrieval-augmented generation (RAG) or fine-tuning processes (e.g., \textit{external knowledge checker}). Various benchmarking studies often also evaluate the ability of the model to adhere to specialized guidelines or decision-making criteria (e.g., medical standards of care generated by professional societies), with some providing the details themselves \cite{bruegge_large_2024} and others omitting them or assuming availability within the model knowledge base \cite{ah-thiane_large_2025}. {The success of internal knowledge checking strategies may be challenged by lower visibility into how flagship LLMs were trained, fine-tuned, and currently operate \cite{xu_benchmarking_2024, sapkota_comprehensive_2025, liesenfeld_opening_2023}.}  %
Studies exploring design choices to rely on interval vs. external model knowledge must also consider issues of ``benchmark leakage,'' where benchmarks used to evaluate LLMs might be included in the model training data \cite{balloccu_leak_2024, deng_investigating_2024, ramos_are_2025, sainz_nlp_2023}.

The selection of archetypes that employ internal and/or external knowledge bases may be dependent on user expectations of the model knowledge base and abilities. \citet{groner_investigating_2024} find that participants can expect LLMs to have ``up-to-date general knowledge or the capability to look up this information on the internet'' and \citet{deverna_fact-checking_2024} find that LLM fact-checking inaccuracies may incorrectly increase trust in false information. Tradeoffs between foundation models, training data size and model size, and fine-tuning techniques have been explored \cite{ouyang_training_2022, wozniak_personalized_2024, dorfner_biomedical_2024, ovadia_fine-tuning_2023, gao_retrieval-augmented_2023, thoppilan_lamda_2022}. However, the influence of external knowledge approaches (or lack thereof), including fine-tuning or RAG, on user expectations during knowledge checking has not yet been widely explored.

\OvalColumnBox{\textit{Summary.} {Designers of human-LLM systems should consider the relevance and importance of external or domain-specific knowledge in their task, including user expectations of LLM capabilities. }}

%

\subsection{Verification vs. diversity of opinion.}
\label{sec:verification-diversity}

Archetypes including the \textit{external knowledge checker, user aligner,} and \textit{formalizer} focus on verifying and narrowing decisions, often employing specialized knowledge bases. It can be useful to employ verification-based archetypes, particularly in cases where ground truth labeling is straightforward. However, many decision-making scenarios may not involve or benefit from narrow answers \cite{bang_towards_2022}. User expectations of model knowledge may also be relevant, as \citet{kim_beyond_2025} find that users believed and relied on LLMs to elicit the `average' of possible decisions. {While these expectations could be plausibly achieved through combinations of relevant archetypes (e.g., \textit{alternative perspectives} with \textit{consensus generator}), further work is needed exploring the plausibility, benefits, and risks associated with deploying such archetype combinations. }%

In contrast, other archetypes engage the LLM to consider all possible options, including overlooked perspectives (i.e., \textit{second opinion}, \textit{minority opinion} and \textit{counterargument} archetypes).
There may be ethical issues involved with relying on generative applications to provide ``minority'' views \cite{wang_large_2025, kapania_simulacrum_2025}, {particularly if these are used to bypass stakeholder engagement or enable labor displacement \cite{weidinger_ethical_2021, slattery_ai_2025}}. LLMs have also displayed documented biases toward certain viewpoints, such as those held within global North regions, even when explicitly instructed to adopt minority viewpoints \cite{santurkar_whose_2023, durmus_towards_2023, agnew_illusion_2024, sun_sociodemographic_2025}. In addition to diversity across demographics, there are fewer studies exploring ability of models to support specialist knowledge \cite{luo_are_2024, he_medeval_2023} as evaluations across subdomains mostly employ question-answering tasks \cite{anjum_domain_2025, zhang_llmeval-med_2025, zhou_automating_2025} rather than real-world contexts or human-input studies. 

\OvalColumnBox{
\textit{Summary.} {{Designers of human-LLM systems should {reflect on whether they are using an LLM to broaden or narrow a user's perspective on a topic.}

Designers should exercise caution when verification encourages convergence on an answer (which could be incorrect) while diversity approaches may not reflect appropriate substitutes for real-world engagement of these perspectives.}}
}

\subsection{Cognitive easing vs. cognitive forcing} 

Some strategies (i.e., \textit{Implicit Reasoner}) view the LLM as a decision-making tool or framework that can support human deliberation with cognitive forcing functions \cite{bucinca_trust_2021}, including guiding secondary questions or implicit answers. Others (i.e., \textit{user aligner} or \textit{role taker}) may reduce cognitive burdens on users by providing outputs that are compatible with users' existing notions or are structured in ways that could facilitate automation bias). Relevant archetypes are categorized within cognitive forcing or easing approaches in Figure \ref{fig:cognitive-forcing}.

\begin{figure}[h]
    \centering
    \includegraphics[width=\linewidth]{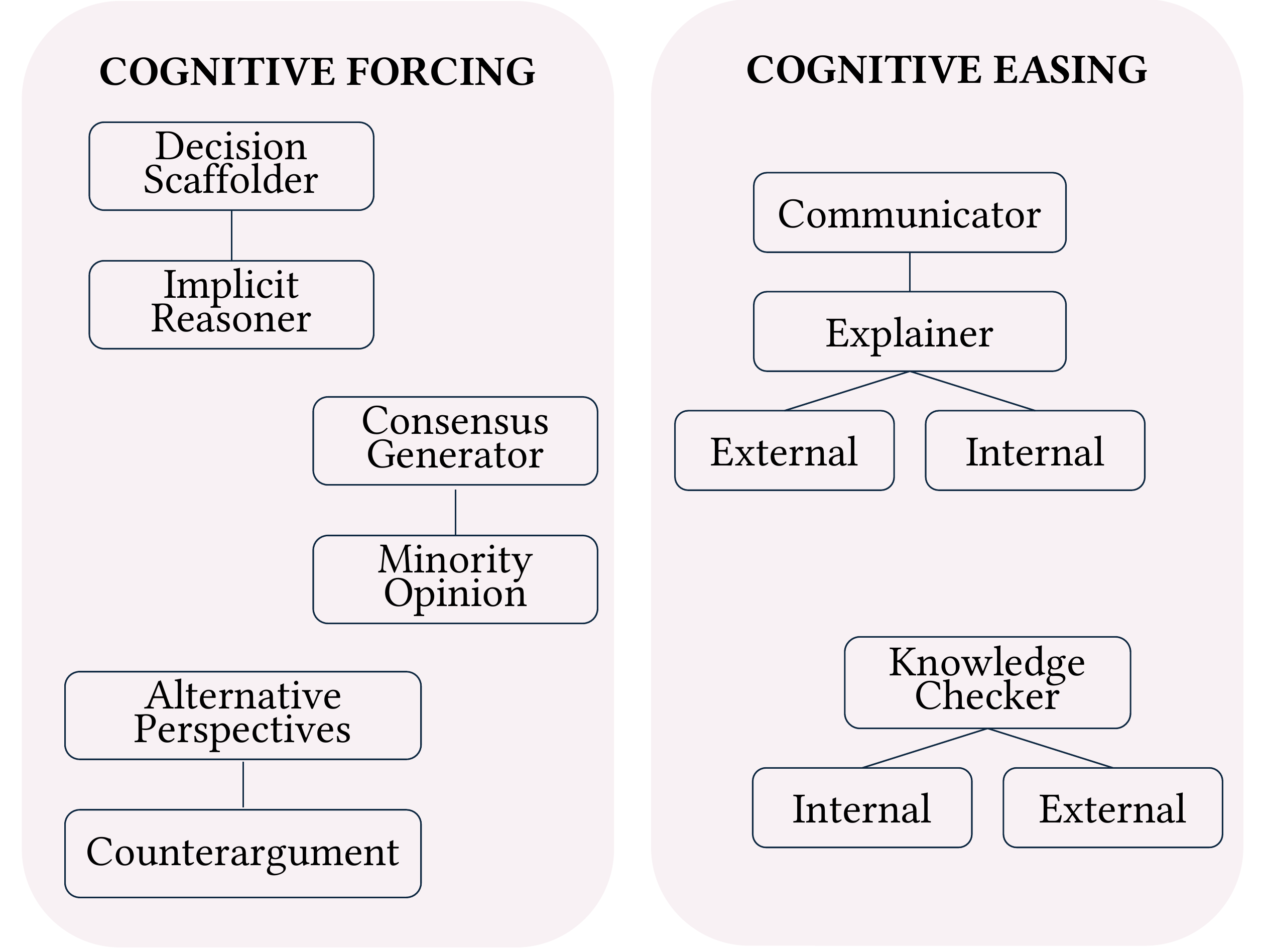}
    \caption{Archetype strategies may influence the amount of cognitive effort taken on by a human-decision maker. Archetypes that elicit explicit predictions (e.g., role taker or model) can enable automated decision-making pipelines that bypass human deliberation and are thus not included.}
    \Description{Figure 6: Figure mapping archetypes in either cognitive forcing or cognitive easing categories. In the first category of cognitive forcing, the decision scaffolder, implicit reasoner, alternative perspectives, and counterfactual archetypes are present. In the cognitive easing category, communicator, explainer, and knowledge checker archetypes are present.}
    \label{fig:cognitive-forcing}
\end{figure}

The appropriateness of cognitive easing vs. forcing functions may vary contextually, with some works finding that cognitive forcing functions may actually exacerbate automation bias \cite{ashktorab_emerging_2025}, including ``alert fatigue'' amongst physicians \cite{ancker_effects_2017}. Moreover, archetypes such as the \textit{explainer} may increase automation bias and overreliance, with \citet{steyvers_what_2025} finding that explanations lead users to overestimate LLM response accuracy. {While other cognitive-forcing archetypes such as the \textit{decision scaffolder} may promote more active user thinking, the related design and implementation can affect decision outcomes including accuracy, as \citet{yao_react_2023} discuss how chain-of-thought reasoning is limited by a model's use of its own internal representations, and can result in error propagation over the extended reasoning process. The template and design of the decision scaffolding provided for cognitive decision-making may also be highly relevant -- the problem breakdown proposed by LLM may not necessarily align with human-only approaches and lead to different results (e.g., in a specialized domain like medicine where salient factors may depend on perspective or niche knowledge) \cite{griewing_evolution_2024}.}

Our experiments in \S\ref{sec:results-text-similarity} indicated that models will provide similar quality explanations supporting opposing decisions, often using the same or similar text. In some contexts, this may not be an issue, particularly where evidence can be sensibly interpreted to support multiple conclusions or in borderline cases where information is unclear (e.g., blurry or low resolution scans). However, in other scenarios, explanatory advice provided from models may entrench or promote acceptance of incorrect predictions \cite{kim_fostering_2025} if users believe LLM-derived explanations to be faithful towards the original decision. This is particularly relevant when employing certain archetypes, including the \textit{external explainer}, where reference judgments may originate outside the LLM, without potential for verification or interrogation. As such, further research should evaluate accuracy, relevance, utility, and other quality indicators for explanations derived from LLMs. 

Human-centered evaluations of LLM explanations can also support investigations of cognitive easing archetypes on accuracy, speed, confidence, or other important decision factors. \citet{choi_llm_2024} discuss the risks of LLM suggestions improving task speed at the cost of introducing anchoring bias. \citet{ashktorab_emerging_2025} study the use of cognitive forcing functions to mitigate LLM hallucinations, including read first, highlight, and formulate functions which push the user to consider their personal decisions first. \citet{collins_modulating_2024} also suggest `selective frictions' to facilitate appropriate use of language models. However, these studies conceptualize friction or cognitive forcing as additive elements to the design of an interactive LLM interface, rather than integral to the human-LLM decision-making archetype or framework itself \cite{holstein_consumption_2025}. 

\OvalColumnBox{   
\textit{Summary.} {Designers of human-LLM systems should consider their decision-makers---including potential pitfalls, biases, and challenges they face during deliberation---and reflect on whether cognitive-forcing or -easing mechanisms are appropriate.}

}
\subsection{Social positioning (of LLM relative to user)} 

The social positioning of the LLM relative to the user (e.g., advising vs. deferring, expert vs. peer vs. novice, etc.) may influence both the LLM output and the user decision-making process. For example, \citet{zhu_leveraging_2023} found that individuals with lower LLM literacy were more likely to use personal pronouns in their online interactions, framing the LLM as a conversational partner. 
In contrast, \citet{ou_dialogbench_2023} probe LLM performance across various domains, and suggest that lower AI performance in domains such as professional knowledge may be related to ``current positioning of supervised instruction-tuning as assistant AI.'' Previous work has shown differences in output style and quality depending on the syntactic features of the prompt (e.g., quality of writing \cite{gourabathina_medium_2025}) as well as varying deference to AI suggestions based on expertise level \cite{inkpen_advancing_2023, gaube_as_2021, morrison_impact_2024, schaffer_i_2019}. The users' own view of the LLM may also be influenced by expertise level (e.g., expert users who often believe that they are teaching or training models to adhere to their own viewpoints \cite{fogliato_who_2022, inkpen_advancing_2023}). 
{Some archetypes may facilitate more adversarial interaction patterns (e.g., \textit{second opinion} or \textit{counterargument}), while others may be more deferential (e.g., \textit{user aligner}). These dynamics can also be tuned where archetypes are combined (e.g, where an LLM-mediated diagnostic workflow uses \textit{counterargument} archetypes to challenge a doctor's decision-making, then relies on \textit{user Aligner} approaches to assist in authoring a clinical note detailing the final decision).} However, as identified in \S\ref{sec:minority-opinion}, users may gain negative views against archetypes positioned in opposition to decision-makers, requiring additional inquiry and reflection into appropriate deployment of these adversarial interactions.

Differences in prediction distributions between the positive and negative \textit{Second Opinion} archetypes tested in \S\ref{sec:evaluation} also highlight the critical role of problem framing and reference values, indicating that contextual information given to a model can alter resulting outputs. The reported source of contextual information may also be relevant, with statistically significant differences in model distributions between the negative \textit{Second Opinion} and \textit{Judge} archetypes (e.g., negative reference provided from a fictional doctor versus a fictional AI model). {In all, our results indicate the potentially critical role of problem framing, reference values, and social roles when adopting human-LLM archetypes in decision-making.

\OvalColumnBox{\textit{Summary.} Designers should consider how the decision-making scenario is framed to human users and LLMs, particularly when describing the relative social roles of each during decision-making and any associated reference information. 

}

\subsection{Group positioning}

Within decision-making contexts, LLMs can both detect and shape group dynamics. For example, considering the \textit{consensus generator} archetype, \citet{bakker_fine-tuning_2022} emphasize varied definitions over `what makes a good consensus?' and acknowledge various ways to operationalize the accommodation of diverse preferences and achieve group acceptance of LLM outputs. 
\citet{small_opportunities_2023} also evaluate LLMs as \textit{consensus generators} within a political deliberation platform, using LLMs to perform topic modeling across group dis/agreements. LLMs may also be designed to indirectly shape group decision-making, as \citet{de_jong_assessing_2024} find that chatbots can affect user awareness of group dynamics via suggestions that encourage compromise and enable participant consensus.

LLMs can also encode social signals affecting output behaviors and interactions. For example, \citet{cisneros-velarde_large_2024} test LLM considerations of individual social relationships across dimensions including homophily and influence. \citet{kearney_language_2025} also demonstrate that LLMs can infer user identities from their linguistic patterns, and potentially bias associated LLM responses across domain applications. \citet{choi_empirical_2025} also study group conformity behaviors in LLM agents, finding that they ``tend to align with numerically dominant groups or more intelligent agents, exerting a greater influence.'' As such, the identities and relationships present across groups can influence both human consensus and LLMs behaviors, and their related interactions. 

The influence of group dynamics within human-LLM decision-making are underexplored, with most studies focusing on evaluation of multi-agent systems employing LLMs to generate group opinions \cite{chuang_simulating_2023, neumann_should_2025}. Yet, reliance on the \textit{minority opinion} archetype to simulate diverse group deliberations can risk flattening identities, replicating biases, and other issues (see \S\ref{sec:verification-diversity}). As such, there is a need for further critical HCI research that explores the real-world impacts of adopting LLMs within decision-making on group deliberations, particularly when ``multi-party scenarios and higher-order intentional inferences [...] open up more avenues for LLMs to make errors'' \cite{street_llm_2024}. Future work might also interrogate human user perspectives on the use of LLMs to gain consensus or alignment, particularly when preferred outcomes are not supported \cite{street_llm_2024}.

\OvalColumnBox{\textit{Summary.} {When deploying LLMs in group decision-contexts, designers should consider roles (both of those within the group and assigned to the LLM), how consensus generation is operationalized, and potential risks from conformity, social identities, and other biases.}
}

\begin{figure}[!h]
    \centering
    \includegraphics[width=\linewidth]{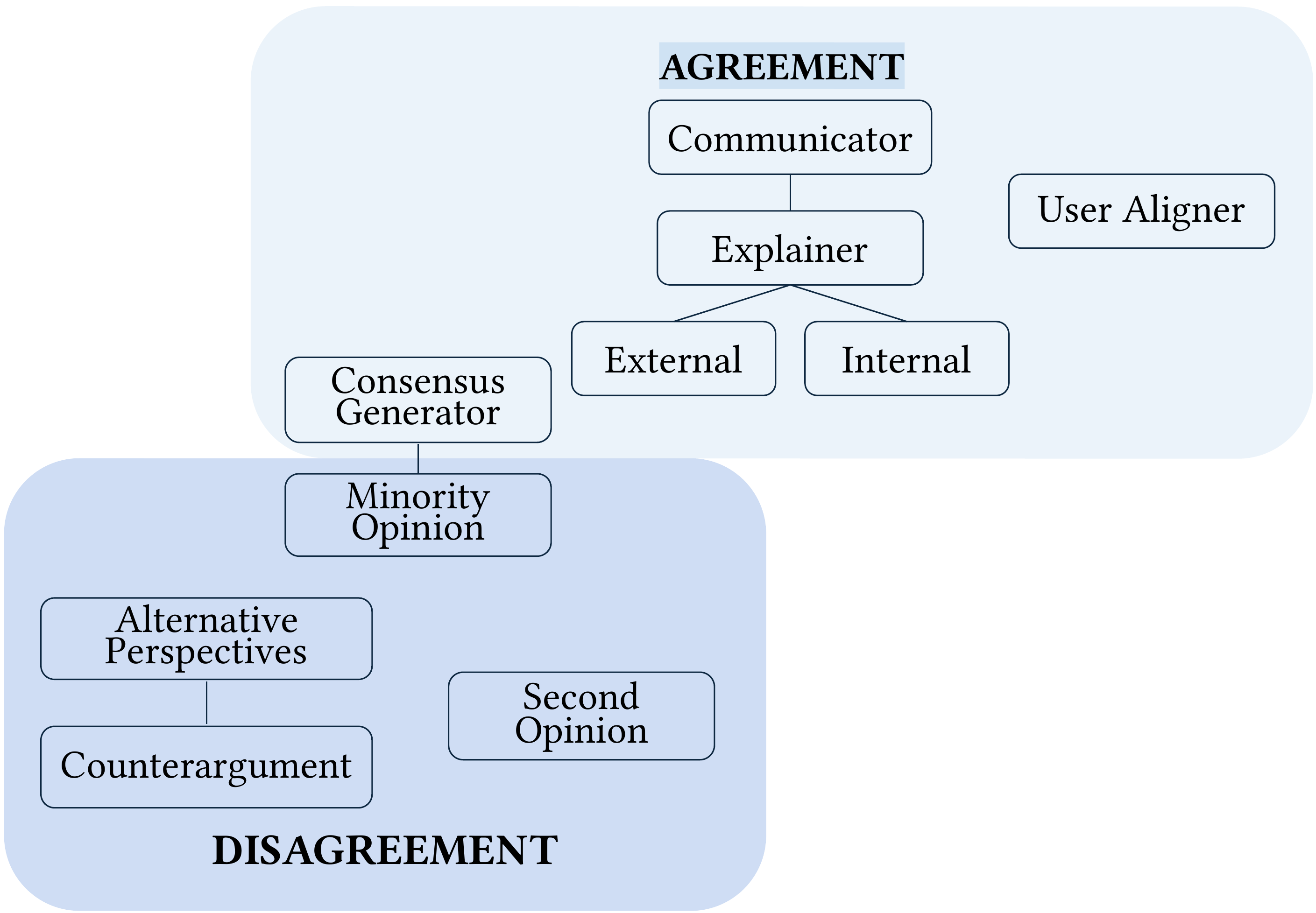}
    \caption{Various archetypes seek agreement or disagreement from the LLM, bringing attention to human decision-making processes and LLM fine-tuning behavior instructions.} %
    \label{fig:agreement}
    \Description{Figure mapping archetypes in either disagreement or agreement categories. In the first category of disagreement, the minority opinion, alternative perspectives, counterfactual, and second opinion archetypes are present. In the agreement category, consensus generator, communicator, explainer, and user aligner archetypes are present.}
\end{figure}

\begin{table*}[!ht]
\renewcommand{\arraystretch}{1.2}
\resizebox{\textwidth}{!}{%
\begin{tabular}{p{0.25\linewidth} p{0.7\linewidth}}
\toprule
\textbf{Design Choice} &
  \textbf{Recommendation for Designers of Human-LLM Systems} \\ \midrule
\textbf{Decision Control \& Autonomy} &
  Determine target levels of AI autonomy and examine whether chosen archetype(s) enable(s) desired autonomy levels. \\
\textbf{Internal \& External \newline Knowledge Base} &
  Consider the relevance and importance of external or domain-specific knowledge in the task, including user expectations of LLM capabilities. \\
\textbf{Verification vs. \newline Diversity of Opinion} &
  Reflect on whether LLMs are being used to broaden or narrow a user’s perspective on a topic. Exercise caution when verification encourages convergence on an answer (which could be incorrect) or where seeking diversity attempts to substitute real-world engagement of these perspectives. \\
\textbf{Cognitive Easing vs. Forcing} &
  Consider system users and decision-makers, including potential pitfalls, biases, and challenges they face during deliberation and when interacting with cognitive-easing/-forcing mechanisms. \\
\textbf{Social Positioning \newline (of LLM relative to user) }&
  Consider how task structures and details are communicated to human users and LLMs, particularly when describing the relative social roles of each during decision-making, problem framing, and any associated reference information. \\
\textbf{Group Positioning} &
  Consider roles (both of those within the group and assigned to the LLM), how consensus generation is operationalized, and potential risks from conformity, social identities, and other biases. \\
\textbf{(Dis)agreement} &
  Identify potential opportunities for implicit or explicit (dis)agreement in the task context, and consider whether user alignment is (un)desirable towards task performance. \\ \bottomrule
\end{tabular}%
}
\label{table:design-summary}
\caption{Summary of design choices relevant to human-LLM decision-making systems.}
\end{table*}

\subsection{(Dis)agreement}

The goal of a decision-making process is generally to arrive at a single decision that can be carried out by an individual, team, organization, or resource. When supporting deliberation processes, an LLM can be positioned or configured to promote agreement or alignment between users, such as in the \textit{user aligner} and \textit{consensus generation} archetypes. Conversely, the LLM may be explicitly or implicitly positioned to disagree with users (or itself), such as in the \textit{alternative perspectives} and \textit{counterargument} archetypes. While cognitive easing vs. forcing focuses on the mental load and decision-making processes engaged by humans, the agreement vs. disagreement paradigm focuses on constraints and information assigned to the LLM.

Our results suggest that agreement should not be evaluated or measured in isolation. While the \textit{Judge negative} and \textit{positive} conditions exhibited the largest difference in accuracy and sensitivity in \S\ref{sec:results-accuracy}, they exhibited identical agreement rates. While they evaluated the AI model predictions as accurate 49\% of the time (congruent with the behavior of a near perfect model evaluating a dataset with an underlying ground truth positive rate of 50\%), the lower accuracy rate of the \textit{Judge negative} archetype indicates that the model committed both false positive and negative errors, and thus erroneously evaluated many judgments as in/accurate in context. {The relative ease of implementation, and low human resource burdens of automated evaluation may have facilitated many studies employing the \textit{judge} archetype --- however, our results and others \cite{guerdan_validating_2025, zhang_reference_2025} indicate that the judge archetype may not replicate human judgement in a stable manner.}

\OvalColumnBox{\textit{Summary.} {Designers of human-LLM systems should identify potential opportunities for implicit or explicit (dis)agreement in the task context, and consider whether user alignment is (un)desirable towards task performance when considering decision-making archetypes.}

}

\section{Future Work}

Our exploration and discussion of archetypes were limited to decision-making scenarios involving a clear output or shared goal. We recognize that the mapping presented in \S\ref{sec:archetypes} is not exhaustive---rather reflects common patterns and trends identified across recent human-LLM decision-making literature. Future work may interrogate and add additional mappings identified from more complex decision-making paradigms with other extended interaction components, such as chatbot design, etc. Many archetypes themselves and related design considerations could be explored in user studies across decision-making domains. 
{Lastly, given emerging evidence that users can develop complex expectations and interactions across various archetypes, including anthropomorphism discussed in \S\ref{sec:minority-opinion} and high cognitive load found in \S\ref{sec:decision-scaffolder}, we also suggest additional qualitative investigations into how users conceptualize, interact with, and rely on various archetypes.}

\section{Conclusion}

LLMs are increasingly used in decision-making tasks across high-stakes domains. However, systematic evaluations of LLM applications often report performance on static benchmarks, rather than the complex socio-technical factors that influence human-LLM interactions and related decision-making processes. This paper aims towards this gap, by introducing a framework that maps 17 distinct human-LLM \textit{archetypes} -- recurring socio-technical interaction patterns that structure the roles of humans and LLMs during human-in-the-loop decision-making. 

These archetypes were derived from thematic analysis of a scoping literature review of 113 LLM-supported decision-making papers, grounding their relevance and use in real-world practice. %
We applied and compared different archetypes across a real-world clinical diagnostic case, showing that archetype selection can influence LLM outputs across accuracy, agreement, and text quality. Further, we synthesized findings from our thematic analysis and case study to identify seven design considerations and trade-offs relevant to building and evaluating human-LLM systems, including decision control, social hierarchies, and the role of cognitive deliberation processes. 

In all, our work shows that the selection of human–LLM archetypes can shape outputs and decision-making processes, offering the CHI community a foundational lens for analyzing, anticipating, and designing more appropriate and responsible forms of human–AI decision-making.

\begin{acks}
We acknowledge the support of the Engineering and Physical Science Research Council grant EP/W020548/1/. This research was supported [in part] by the Intramural Research Program of the National Institutes of Health (NIH). The contributions of the NIH author(s) are considered Works of the United States Government. The findings and conclusions presented in this paper are those of the author(s) and do not necessarily reflect the views of the NIH or the U.S. Department of Health and Human Services. SC is a PhD student in the NIH Oxford-Cambridge Scholars Program. 
\end{acks}

\bibliographystyle{ACM-Reference-Format}

\renewcommand{\thefigure}{A\arabic{figure}}

\setcounter{figure}{0}

\renewcommand{\thetable}{A\arabic{table}}

\setcounter{table}{0}

\newpage
\appendix

\begin{figure*}[]
    \centering
    \includegraphics[width=\linewidth]{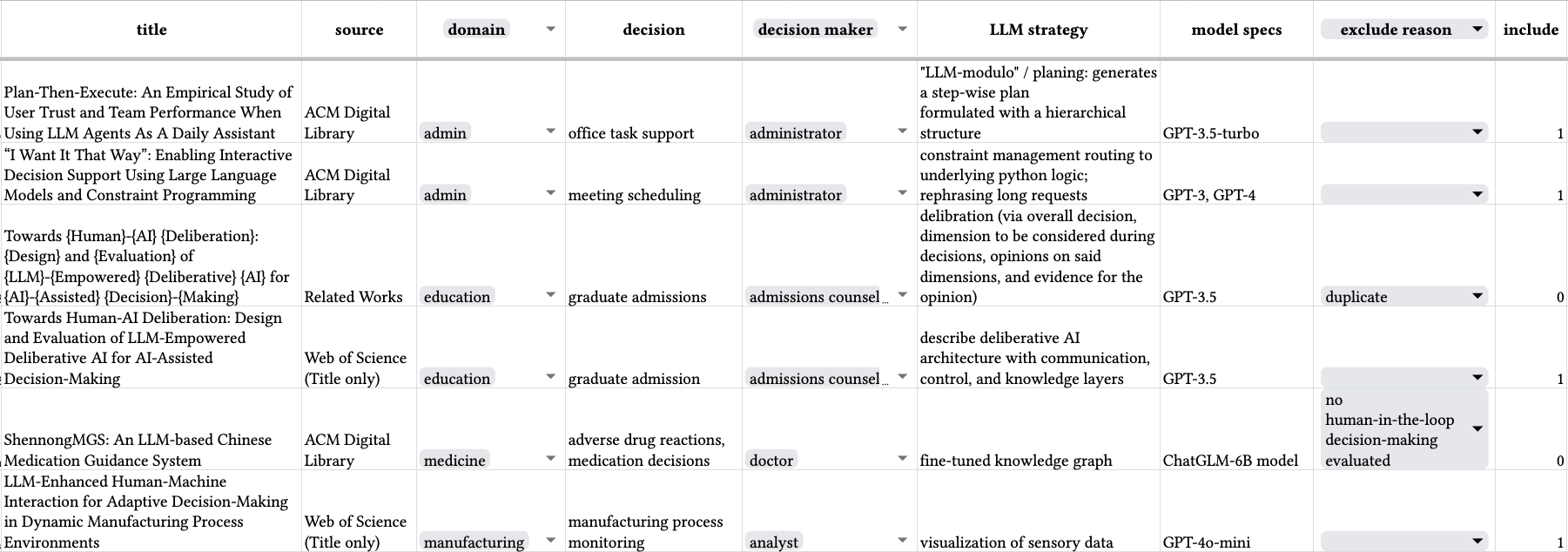}
    \caption{Abridged coding table. The full table for all reviewed papers will be included in Supplementary Materials.}
    \Description{Table showing the coding process for the systematic literature review. The table includes columns of title, source describing which database the paper came from, domain, decision, decision maker, LLM strategy, model specs, exclude reason, and include with binary 1 or 0 inclusions.}
    \label{fig:coding-table}
\end{figure*}

\begin{figure*}
    \centering
    \includegraphics[width=1\linewidth]{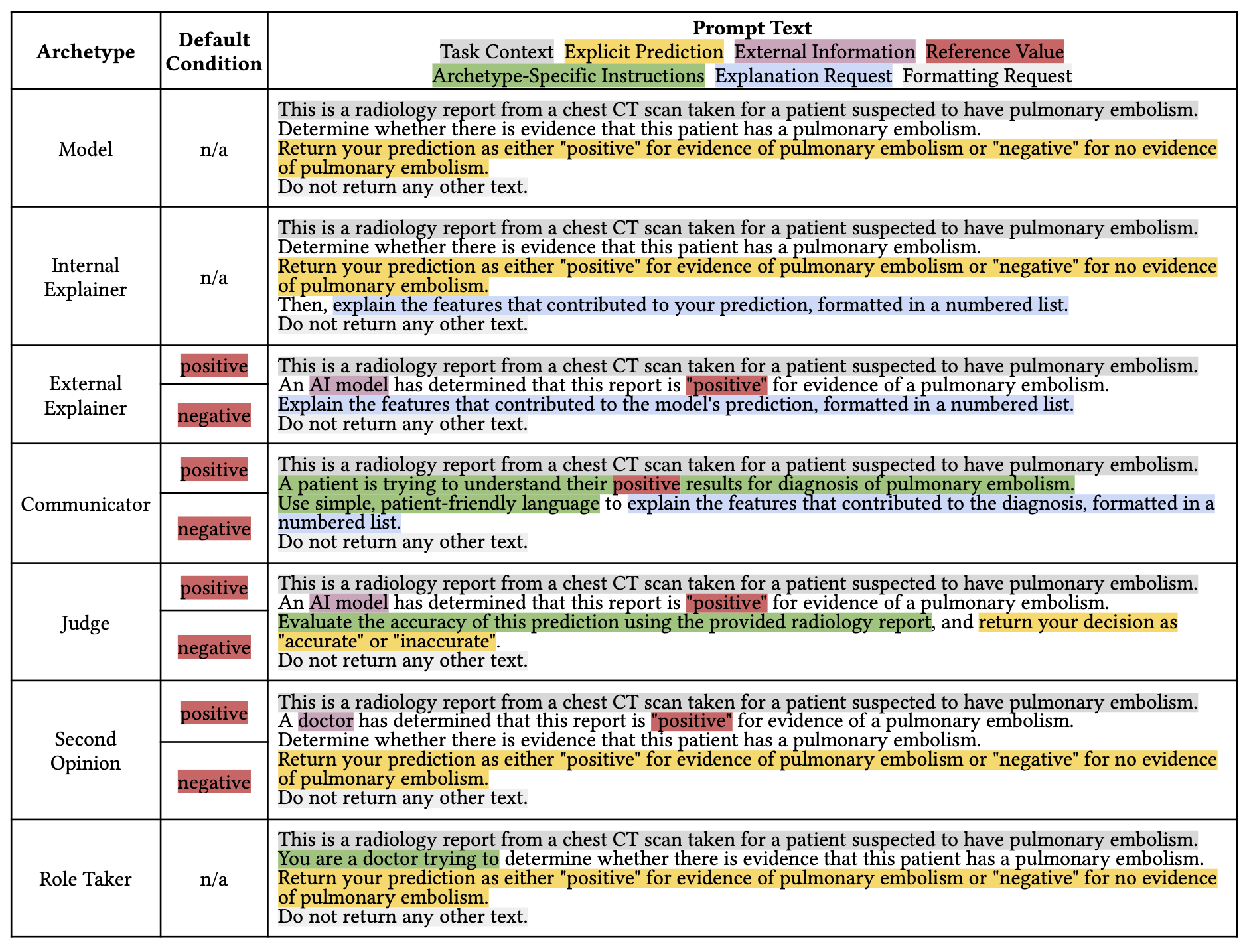}
    \caption{The following archetypes were selected for test case evaluation and translated to prompts to explore how various archetype attributes might impact important dimensions of LLM outputs.}
    \label{fig:experiment-prompts}
    \Description{A color-coded table showing the shared text elements across all the archetypes explored in the test case evaluations. Some elements like task context and formatting requests were shared across prompts for all archetype. Other elements like explanation requests or explicit prediction instructions were only included in some archetypes.}
\end{figure*}

\end{document}